\documentclass[12pt]{article}
\usepackage{graphicx}
\begin{document}
\begin{center}
{\large\bf Gravitational Theory, Galaxy Rotation Curves and
Cosmology without Dark Matter} \vskip 0.3 true in {\large J. W.
Moffat} \vskip 0.3 true in {\it The Perimeter Institute for
Theoretical Physics, Waterloo, Ontario, N2J 2W9, Canada} \vskip
0.3 true in and \vskip 0.3 true in {\it Department of Physics,
University of Waterloo, Waterloo, Ontario N2Y 2L5, Canada}
\end{center}
\begin{abstract}%
Einstein gravity coupled to a massive skew symmetric field
$F_{\mu\nu\lambda}$ leads to an acceleration law that modifies the
Newtonian law of attraction between particles. We use a framework
of non-perturbative renormalization group equations as well as
observational input to characterize special renormalization group
trajectories to allow for the running of the effective
gravitational coupling $G$ and the coupling of the skew field to
matter. Strong renormalization effects occur at large and small
momentum scales. The latter lead to an increase of Newton's
constant at large galactic and cosmological distances. For weak
fields a fit to the flat rotation curves of galaxies is obtained
in terms of the mass (mass-to-light ratio $M/L$) of galaxies. The
fits assume that the galaxies are not dominated by exotic dark
matter and that the effective gravitational constant $G$ runs with
distance scale. The equations of motion for test particles yield
predictions for the solar system and the binary pulsar PSR 1913+16
that agree with the observations. The gravitational lensing of
clusters of galaxies can be explained without exotic dark matter.
An FLRW cosmological model with an effective $G=G(t)$ running with
time can lead to consistent fits to cosmological data without
assuming the existence of exotic cold dark matter.
\end{abstract}
\vskip 0.2 true in e-mail: jmoffat@perimeterinstitute.ca


\section{Introduction}

A nonsymmetric gravity theory (NGT) has been applied to explain
rotation curves of galaxies and cosmology, without invoking
dominant dark matter and identifying dark energy with the
cosmological constant~\cite{Moffat,Moffat2,Moffat3,Moffat4}. In
the following, we develop a simpler gravitational theory based on
Einstein's general relativity (GR) and a skew symmetric rank three
tensor field $F_{\mu\nu\lambda}$, forming a
metric-skew-tensor-gravity (MSTG) theory. We shall apply this
gravity theory to explain the flat rotation curves of galaxies and
cluster lensing without postulating exotic dark matter. The weak
field approximation of the NGT field equations corresponds to the
equivalent field equations in MSTG and, therefore, to this order
of approximation will yield the same predictions as MSTG. The more
complicated NGT will have different possible consequences for
strong gravitational fields. However, in the following, we shall
be concerned with the observational data for the solar system, the
binary pulsar PSR 1913+16, galaxies, clusters of galaxies and
cosmology. The MSTG theory provides a more direct way of deriving
a comparison with the data.

Since no dark matter has been detected so far, it seems imperative
to seek a possible modified gravitational theory that could
explain the now large amount of data on galaxy rotation curves. A
cosmological model obtained from the field equations and a running
of the effective gravitational coupling constant $G$ in the matter
epoch could explain the growth of large scale structure formation
without invoking cold dark matter. The running of the cosmological
constant would produce a quintessence-like dark energy that could
account for the acceleration of the expansion of the
universe~\cite{Perlmutter,Riess,Spergel}.

A renormalization group (RG) framework ~\cite{Reuter} for MSTG is
developed to describe the running of the effective gravitational
coupling constant $G$, and the effective coupling constant
$\gamma_c$ that measures the strength of the coupling of the
$F_{\mu\nu\lambda}$ field to matter. A momentum cutoff
identification $k=k(x)$ associates the RG scales to points in
spacetime. For the weak field static, spherically symmetric
solution the RG flow equations allow a running with momentum $k$
and distance $\ell=1/k$ for the effective Newtonian coupling
constant $G=G(r)$, the coupling constant $\gamma_c=\gamma_c(r)$,
and the effective mass of the skew field $\mu=\mu(r)$ where $r$
denotes the radial coordinate. The form of $G(r)$ as a function of
$r$, obtained from the modified Newtonian acceleration law, leads
to agreement with solar system observations, terrestrial
gravitational experiments and the binary pulsar PSR 1913+16
observations, while strong renormalization effects in the infrared
regime at large distances lead to fits to galaxy rotation curves.

A fit to both low surface brightness and high surface brightness
galaxies is achieved in terms of the total galaxy mass $M$ (or
$M/L$) without exotic dark matter. A satisfactory fit is achieved
to the rotational velocity data generic to the elliptical galaxy
NGC 3379. Fits to the data of the two spheroidal dwarf galaxies
Fornax and Draco and the globular cluster $\omega$ Centauri are
also obtained. The predicted light bending and lensing can lead to
agreement with galaxy cluster lensing observations. A model of the
modified acceleration law that includes a description of radial
velocity curves in the core of galaxies as well as in the outer
regions of the galaxy is shown to yield good fits to rotational
velocity data.

For a homogeneous and isotropic universe, based on the Friedmann-
Lemaitre-Robertson-Walker (FLRW) model, the average value of the
skew symmetric field $F_{\mu\nu\lambda}$ is zero. However, for an
effective gravitational constant $G=G(t)$ that runs with time, we
find that the constraint $G(t)\sim G_0$ (where $G_0$ is Newton's
constant, $G_0=6.673\times 10^{-8}\,{\rm g}^{-1}\,{\rm cm}^3\,{\rm
sec}^{-2}$) must be imposed on $G(t)$ at the time of big bang
nucleosynthesis (BBN) and the surface of last scattering
(decoupling) when the fraction of baryons is $\Omega_B\sim 0.04$.
For a rapid increase of $G(t)$ after the time of decoupling when
$G(t)$ reaches a value $G(t)\sim 6G_0$, the baryonic matter
fraction becomes $\Omega_M\sim 6\Omega_B\sim 0.24$ and there is no
need for undetected non-baryonic dark matter\footnote{An
alternative cosmological model has been proposed~\cite{Blanchard}
that fits the data without a cosmological constant and for a
universe consisting only of baryons and cold dark matter and with
a low value of the Hubble constant: $H_0\sim 46\,{\rm km}/s/{\rm
Mpc}$.}

\section{The Field Equations}

The action for the MSTG theory is given by
\begin{equation}
S=S_G+S_F+S_{FM}+S_M,
\end{equation}
where $S_G$ is the Einstein-Hilbert action (we choose units such
that $c=1$):
\begin{equation}
\label{gravaction} S_G=\frac{1}{16\pi G}\int
d^4x\sqrt{-g}(R-2\Lambda),
\end{equation}
where $g={\rm Det}(g_{\mu\nu})$, $g_{\mu\nu}$ is the symmetric
metric tensor of pseudo-Riemannian geometry,
$R=g^{\mu\nu}R_{\mu\nu}$ is the Ricci scalar and $\Lambda$ is the
cosmological constant. Moreover, $S_F$ is the skew field action
\begin{equation}
\label{skewaction} S_F=\int
d^4x\sqrt{-g}(\frac{1}{12}F_{\mu\nu\rho}F^{\mu\nu\rho}-\frac{1}{4}\mu^2A_{\mu\nu}A^{\mu\nu}),
\end{equation}
where
\begin{equation}
F_{\mu\nu\lambda}=\partial_\mu A_{\nu\lambda}+\partial_\nu
A_{\lambda\mu}+\partial_\lambda A_{\mu\nu},
\end{equation}
and $\mu$ is the mass of the skew field. The gravitational
constant $G$ in the action $S_G$ is defined in terms of the
``bare'' gravitational constant $G_0$:
\begin{equation}
\label{gravirenorm} G=G_0Z,
\end{equation}
where $Z$ corresponds to a ``renormalization'' of $G$.

The actions $S_M$ and $S_F$ satisfy the relations
\begin{equation}
\frac{1}{\sqrt{-g}}\frac{\delta S_M}{\delta
g^{\mu\nu}}=-\frac{1}{2}T_{M\mu\nu},\quad\frac{1}{\sqrt{-g}}\frac{\delta
S_F}{\delta g^{\mu\nu}}=-\frac{1}{2}T_{F\mu\nu}.
\end{equation}
Moreover, we have
\begin{equation}
\quad\frac{\delta S_{FM}}{\delta A^{\mu\nu}}=-J_{\mu\nu}.
\end{equation}
Here, $T_{M\mu\nu}$ is the energy-momentum tensor for matter,
while $T_{F\mu\nu}$ is the energy-momentum tensor containing the
contributions from the massive skew field $A_{\mu\nu}$. Also,
$J_{\mu\nu}$ is the tensor density source for the $A_{\mu\nu}$
field.

A possible action for the skew field matter coupling
is~\cite{Damour}:
\begin{equation}
\label{Faction} S_{FM}=\int
d^4xF_{\lambda\mu\nu}J^{*\lambda\mu\nu} =-3\int
d^4x\epsilon^{\alpha\beta\mu\nu}A_{\alpha\beta}\partial_\mu J_\nu,
\end{equation}
where $J_\mu$ is a vector current that can be conserved and
$J^{*\mu\nu\lambda}=\epsilon^{\mu\nu\lambda\alpha}J_\alpha$ is the
dual tensor current density. The Levi-Civita symbol
$\epsilon^{\mu\nu\lambda\alpha}$ transforms as a contravariant
tensor density, while $\epsilon^{\mu\nu\lambda\alpha}=\pm 1, 0$.
The Pauli coupling (\ref{Faction}) is gauge invariant for any
$J_\mu$. The source current can be identified with a matter
fermion current associated with baryon and lepton number.

The field equations derived from the action principle are given by
\begin{equation}
\label{Gequation} G_{\mu\nu}+\Lambda g_{\mu\nu}=8\pi GT_{\mu\nu},
\end{equation}
\begin{equation}
\label{Ffieldequations} \nabla^\sigma
F_{\mu\nu\sigma}+\mu^2A_{\mu\nu}=\frac{1}{\sqrt{-g}}J_{\mu\nu},
\end{equation}
where $G_{\mu\nu}=R_{\mu\nu}-\frac{1}{2}g_{\mu\nu}R$,
$T_{\mu\nu}=T_{M\mu\nu}+T_{F\mu\nu}$ and $\nabla^\sigma$ denotes
the covariant derivative with respect to $g_{\mu\nu}$. Moreover,
we have
\begin{equation}
J_{\mu\nu}=\epsilon_{\mu\nu\alpha\beta}\partial^\alpha J^\beta.
\end{equation}

The Bianchi identities satisfied by the Einstein tensor
$G^{\mu\nu}$ lead to the conservation laws
\begin{equation}
\nabla_\nu T^{\mu\nu}=0.
\end{equation}

\section{Equations of Motion of Test Particles}

The equations of motion for a test particle are given by
\begin{equation}
\label{geodesic2} \frac{du^\mu}{d\tau}+ \left\{{\mu\atop
\alpha\beta}\right\}u^\alpha u^\beta
=g^{\mu\alpha}f_{\alpha\nu}u^\nu,
\end{equation}
where $\tau$ is the proper time along the path of the particle and
$u^\lambda=dx^\lambda/d\tau$ is the 4-velocity of the particle.
Moreover,
\begin{equation}
\left\{{\lambda\atop \mu\nu}\right\}={1\over 2}g^{\lambda\rho}
\left(g_{\mu\rho,\nu}+g_{\rho\nu,\mu}-g_{\mu\nu,\rho}\right),
\end{equation}
is the Christoffel connection and
\begin{equation}
f_{\alpha\mu}=\lambda\partial_{[\alpha}
\biggl(\frac{\epsilon^{\eta\sigma\nu\lambda}}{\sqrt{-g}}
H_{\sigma\nu\lambda}g_{\mu]\eta}\biggr).
\end{equation}
Here, $H_{\mu\nu\lambda}$ is given by
\begin{equation}
H_{\mu\nu\lambda}=\frac{1}{3}(\partial_\lambda A_{\mu\nu}
+\partial_\mu A_{\nu\lambda}+\partial_\nu A_{\lambda\mu}),
\end{equation}
and $\lambda$ is a coupling constant with the dimension of length
that couples the skew field to the test particle.

For a spherically symmetric, static skew symmetric potential field
$A_{\mu\nu}$ there are two non-vanishing components, the
``magnetic'' field potential $A_{0r}(r)=w(r)$ and the ``electric''
potential field $A_{\theta\phi}(r)=f(r)\sin\theta$. We shall
assume that only the electric field contribution $f(r)\sin\theta$
is non-zero (no static magnetic poles). Then, the tensor
$H_{\mu\nu\lambda}$ has only one non-vanishing component:
\begin{equation}
H_{\theta\phi r}=\frac{1}{3}\partial_rA_{\theta\phi}=f'\sin\theta,
\end{equation}
where $f'=df/dr$. We obtain
\begin{equation}
f_{r0}=\lambda\frac{d}{dr}\biggl(\frac{\gamma
f^\prime}{\sqrt{\alpha\gamma r^4}}\biggr).
\end{equation}

For a static spherically symmetric gravitational field the line
element is
\begin{equation}
\label{lineelement}
ds^2=\gamma(r)dt^2-\alpha(r)dr^2-r^2(d\theta^2+\sin^2\theta
d\phi^2).
\end{equation}
The equations of motion for a test particle are given by
\begin{equation}
\label{rmotion}
\frac{d^2r}{d\tau^2}+\frac{\alpha'}{2\alpha}\biggl(\frac{dr}{d\tau}\biggr)^2
-\frac{r}{\alpha}\biggl(\frac{d\theta}{d
\tau}\biggr)^2-r\biggl(\frac{\sin^2\theta}{\alpha}\biggr)\biggl(\frac{d\phi}{d\tau}\biggr)^2
+\frac{\gamma'}{2\alpha}\biggl(\frac{dt}{d\tau}\biggr)^2
$$ $$
+\frac{1}{\alpha}\frac{d}{dr}\biggl(\frac{\lambda\gamma
f'}{\sqrt{\alpha\gamma
r^4}}\biggr)\biggl(\frac{dt}{d\tau}\biggr)=0,
\end{equation}
\begin{equation}
\label{tequation}
\frac{d^2t}{d\tau^2}+\frac{\gamma'}{\gamma}\biggl(\frac{dt}{d\tau}\biggr)
\biggl(\frac{dr}{d\tau}\biggr)+\frac{1}{\gamma}\frac{d}{dr}\biggl(\frac{\lambda\gamma
f'}{\sqrt{\alpha\gamma
r^4}}\biggr)\biggl(\frac{dr}{d\tau}\biggr)=0,
\end{equation}
\begin{equation}
\frac{d^2\theta}{d\tau^2}+\frac{2}{r}\biggl(\frac{d\theta}{d\tau}
\biggr)\biggl(\frac{dr}{d\tau}\biggr)-\sin\theta\cos\theta\biggl(\frac{d\phi}{d\tau}\biggr)^2
=0,
\end{equation}
\begin{equation}
\label{phiequation}
\frac{d^2\phi}{d\tau^2}+\frac{2}{r}\biggl(\frac{d\phi}{d\tau}\biggr)\biggl(\frac{dr}{d\tau}\biggr)
+2\cot\theta\biggl(\frac{d\phi}{d\tau}\biggr)\biggl(\frac{d\theta}{d\tau}\biggr)=0.
\end{equation}

The orbit of the test particle can be shown to lie in a plane and
by an appropriate choice of axes, we can make $\theta=\pi/2$.
Integrating Eq.(\ref{phiequation}) gives
\begin{equation}
\label{angular} r^2\frac{d\phi}{d\tau}=J,
\end{equation}
where $J$ is the conserved orbital angular momentum. Integration
of Eq.(\ref{tequation}) gives
\begin{equation}
\label{dtequation}
\frac{dt}{d\tau}=-\frac{1}{\gamma}\biggl[\frac{\lambda\gamma
f'}{\sqrt{\alpha\gamma r^4}}+E\biggr],
\end{equation}
where $E$ is the constant energy per unit mass.

By substituting (\ref{dtequation}) into (\ref{rmotion}) and using
(\ref{angular}), we obtain
\begin{equation}
\label{reducedrmotion}
\frac{d^2r}{d\tau^2}+\frac{\alpha'}{2\alpha}\biggl(\frac{dr}{d\tau}\biggr)^2
-\frac{J^2}{\alpha
r^3}+\frac{\gamma'}{2\alpha\gamma^2}\biggl(\frac{\lambda\gamma
f'}{\sqrt{\alpha\gamma r^4}}+E\biggr)^2
=\frac{1}{\alpha\gamma}\frac{d}{dr}\biggl(\frac{\lambda\gamma
f'}{\sqrt{\alpha\gamma r^4}}\biggr)\biggl(\frac{\lambda\gamma
f'}{\sqrt{\alpha\gamma r^4}}+E\biggr).
\end{equation}

We do not have an exact spherically symmetric, static solution for
massive MSTG. For large enough values of $r$, we shall approximate
the metric components $\alpha$ and $\gamma$ by the Schwarzschild
solution:
\begin{equation}
\label{Schwarzschild} \alpha(r)\sim\frac{1}{1-\frac{2GM}{r}},\quad
\gamma(r)\sim 1-\frac{2GM}{r},
\end{equation}
and make the approximations that $2GM/r\ll 1,\lambda f'/r^2\ll 1$,
$f/r^2\ll 1$ and the slow motion approximation $dr/dt\ll 1$. Then,
for material particles we set $E=1$ and (\ref{reducedrmotion})
becomes
\begin{equation}
\label{Newton} \frac{d^2r}{dt^2}-\frac{J_N^2}{r^3}+\frac{GM}{r^2}
=\lambda\frac{d}{dr}\biggl(\frac{f'}{r^2}\biggr),
\end{equation}
where $J_N$ is the Newtonian orbital angular momentum.

\section{Linear Weak Field Approximation}

We expand $g_{\mu\nu}$ about a Minkowski flat spacetime
\begin{equation}
g_{\mu\nu}=\eta_{\mu\nu}+h_{\mu\nu}+O(h^2),
\end{equation}
where $\eta_{\mu\nu}={\rm diag}(1,-1,-1,-1)$ is the Minkowski
metric tensor. The skew field $F_{\mu\nu\lambda}$ obeys the
linearized equations of motion for empty space
\begin{equation}
\label{linearizedeq}
\partial^\sigma
F_{\mu\nu\sigma}+\mu^2A_{\mu\nu}=0.
\end{equation}
These are the massive Kalb-Ramond-Proca equations~\cite{Ramond},
which are free of ghost pole instabilities and possess a positive
Hamiltonian bounded from below.

For the components $A_{\theta\phi}=f(r)\sin\theta$, we find from
(\ref{linearizedeq}) for a static spherically symmetric field:
\begin{equation}
\label{weakfequation}
 f''(r)-\frac{2}{r}f'(r)-\mu^2 f(r)=0,
\end{equation}
which has the solution
\begin{equation}
\label{fweak} f(r)=\frac{1}{3}sG^2M^2\exp(-\mu r)(1+\mu r),
\end{equation}
where $s$ is a dimensionless constant. We note that
$A_{\theta\phi}$ are the components of a second rank skew tensor,
which are not equivalent for a massive skew field to a pure scalar
field. We obtain (\ref{weakfequation}) when we transform to the
spherically symmetric equations in polar coordinates.

Let us now choose $g_{\mu\nu}$ to be a Ricci-flat GR background
metric $g_{\mu\nu}=g^{GR}_{\mu\nu}$. The skew field
$F_{\mu\nu\sigma}$ obeys the linearized equation of motion in the
GR background geometry
\begin{equation}
\label{GRlinearizedeq} \nabla^\sigma
F_{\mu\nu\sigma}+4A^{\sigma\beta}B_{\beta\mu\sigma\nu}+\mu^2A_{\mu\nu}=0,
\end{equation}
where $\nabla^\lambda$ and $B_{\beta\mu\sigma\nu}$ denote the
background GR covariant derivative and curvature tensor,
respectively.

The skew field components $A_{0r}$ have the
solution~\cite{Clayton}:
\begin{equation}
(2\alpha''-\mu^2)A_{0r}=0,
\end{equation}
where $\gamma(r)\sim 1/\alpha(r)\sim 1-2GM/r$. We conclude from
this that $A_{0r}$ vanishes outside the source. In the massless
case, this is the surviving spherically symmetric ghost mode which
in this case is pure gauge. When the mass term is added, although
these modes are no longer pure gauge, they do not propagate since
they are locally coupled to the source. Thus, in the absence of
sources only the $A_{\theta\phi}$ components are physically
interesting.

For the static, spherically symmetric spacetime, the linearized
equation of motion for $f(r)$ on a Schwarzschild background takes
the form
\begin{equation}
\biggl(1-\frac{2GM}{r}\biggr)f''-\frac{2}{r}\biggl(1-\frac{3GM}{r}\biggr)f'
-\biggl(\frac{\mu^2}{r^2}+\frac{8GM}{r}\biggr)f=0,
\end{equation}
The solution to this equation in leading order
is~\cite{Moffat4,Clayton,Cornish}:
\begin{equation}
f(r)=\frac{sG^2M^2}{3}\frac{\exp(-\mu r)}{(\mu r)^{\mu
GM}}\biggl(1+\mu r+ \frac{GM}{r}\biggl[2+\mu r\exp(2\mu
r)Ei(1,2\mu r)(\mu r-1)\biggr]\biggr),
\end{equation}
where $Ei$ is the exponential integral function
\begin{equation}
Ei(n,x)=\int_1^{\infty}dt\frac{\exp(-xt)}{t^n}.
\end{equation}

We obtain for large $r$ the solution
\begin{equation}
f(r)=\frac{1}{3}\frac{sG^2M^2\exp(-\mu r)(1+\mu r)}{(\mu r)^{\mu
GM}}.
\end{equation}
For $\mu GM\ll 1$, we have $(\mu r)^{\mu GM}\sim 1$, yielding the
solution (\ref{fweak}).

To the order of weak field approximation, we obtain from
Eqs.(\ref{Newton}) and (\ref{fweak}):
\begin{equation}
\label{Yukawa} \frac{d^2r}{dt^2}
-\frac{J_N^2}{r^3}=-\frac{GM}{r^2}+\frac{\sigma\exp(-\mu
r)}{r^2}(1+\mu r),
\end{equation}
where the constant $\sigma$ is given by
\begin{equation}
\label{sigma} \sigma=\frac{\lambda sG^2M^2\mu^2}{3}.
\end{equation}
We have required in Eq.(\ref{Yukawa}) that the additional
acceleration on the right-hand side is a repulsive force. This is
in keeping with the identification of $A_{\mu\nu}$ with the
potential field of a massive axial vector spin $1^+$ boson.

An analysis of the physical properties of a massive spin $1^+$
boson exchange was carried out by Damour, Deser and McCarthy
(DDM)~\cite{Damour}, based on a modification of a massive NGT with
skew field $B_{\mu\nu}$. In a version of a rigorous massive
NGT~\cite{Moffat3,Moffat4}, the weak field solution for a massive
Kalb-Ramond field ~\cite{Ramond} follows consistently from a weak
field approximation to the rigorous NGT field equations and leads
to a stable vacuum without ghost energy modes. The solution of the
weak field approximation NGT field equations yields the solution
(\ref{fweak}) for a spherically symmetric system.

The macroscopic coupling action for the DDM skew symmetric
$B_{\mu\nu}$ field is given in their notation by
\begin{equation}
S_J=-\frac{1}{6}f_c\int d^4x H_{\lambda\mu\nu}J^{*\lambda\mu\nu}
=\frac{1}{2}f_c\int
d^4x\epsilon^{\mu\nu\alpha\beta}B_{\alpha\beta}\partial_\mu J_\nu,
\end{equation}
where $f_c$ denotes a dimensionless coupling constant,
$H_{\lambda\mu\nu}$ is the fully skew symmetric field strength and
$J^\mu$ is a current vector that can be conserved. Moreover,
$J^{*\lambda\mu\nu}=\epsilon^{\lambda\mu\nu\alpha}J_\alpha$ is a
totally antisymmetric tensor density. The action $S_J$ (within
trivial numerical factors) is the same action (\ref{Faction}) that
we have used to obtain our field equations (\ref{Ffieldequations})
for the $A_{\mu\nu}$ potential field and the equations of motion
of a test particle (\ref{geodesic2}). DDM show that the lowest
order matter coupling of the skew B-field corresponds to a vector
particle ``fifth force'' coupled to the current $J^\mu$, which can
be identified with a fermion current with the dimensionless
coupling $g_5=\sqrt{4\pi G_0}\mu f_c$. Note that the coupling
constant $g_5$ is inversely proportional to the range of the fifth
force.

We can consider the strength of the skew field coupling by
introducing the coupling constant
\begin{equation}
\alpha_c\equiv\frac{f_c}{m_N}\sim 2f_c\times 10^{-14}\,{\rm cm},
\end{equation}
where $m_N$ is the nucleon mass (one atomic mass unit). Moreover,
$\alpha_c$ has dimensions of a length and couples baryon number to
$m_NJ^\mu$.

The coupling constant $\alpha_c$ can be expressed in terms of
$\alpha_5=g_5^2/(4\pi m_N^2)$ as
\begin{equation}
\label{lambdacoupling}
\alpha_c=\frac{\sqrt{\alpha_5}}{\mu}=\sqrt{\alpha_5}r_0,
\end{equation}
where $r_0=1/\mu$ is the range of the force. This show that
$\alpha_c$ can take large macroscopic values provided that the
range $r_0$ is large enough. In our RG flow framework, the
coupling constant $\alpha_c$ will run and become increasingly
large in the IR momentum limit $k\rightarrow 0$, corresponding to
increasing values of the range $r_0$. We can define a
``renormalized'' coupling constant
\begin{equation}
\alpha_c=\alpha_{c0}A,
\end{equation}
where $\alpha_{c0}$ is the bare skew field coupling constant and
$A$ is a renormalization constant.

Experiments have put stringent bounds on possible fifth forces and
the magnitude of $g_5$~\cite{Adelberger}. However, DDM make the
important points for phenomenological purposes that the strength
of the coupling $f_c$ is unbounded as the range increases, and
that the magnitude of the B-field (in our notation the B-field is
$A_{\mu\nu}$) is primarily proportional to $f_c$ and independent
of the range $r_0$. This allows the B-field to have a
``gravitational'' strength, while still keeping compatibility with
the existing stringent bounds on possible violations of the weak
equivalence principle and composition-dependent effects in
Newtonian gravity.

It should be noted that significant observational bounds on fifth
forces only apply to distances $ < 100\, AU$ in the solar system
and outer solar system, corresponding to $\mu \sim 10^{-20}$ eV.
The experiments do not place useful bounds on the strength of the
coupling of a gravitational fifth force at galactic distance
scales $\sim 5-100$ kpc or at cosmological scales. This is in
keeping with the requirement that our modified acceleration law
will be approximately Newtonian at distance scales within the
solar system, and that the MSTG predictions for solar system
distance scales agree with the solar system observations.

\section{Orbital Equation of Motion}

We set $\theta=\pi/2$ in (\ref{lineelement}), divide the resulting
expression by $d\tau^2$ and use Eqs.(\ref{angular}) and
(\ref{dtequation}) to obtain
\begin{equation}
\label{energyconserved}
\biggl(\frac{dr}{d\tau}\biggr)^2+\frac{J^2}{\alpha
r^2}-\frac{1}{\alpha\gamma}\bigg[\frac{\lambda\gamma f'}
{\sqrt{\alpha\gamma r^4}}+E\biggr]^2=-\frac{E}{\alpha}.
\end{equation}
We have $ds^2=Ed\tau^2$, so that $ds/d\tau$ is a constant. For
material particles $E>0$ and for massless photons $E=0$.

Let us set $u=1/r$ and by using (\ref{angular}), we have
$dr/d\tau=-Jdu/d\phi$. Substituting this into
(\ref{energyconserved}), we obtain
\begin{equation}
\label{neworbital} \biggl(\frac{du}{d\phi}\biggr)^2=
\frac{1}{\alpha\gamma J^2}\biggl[E+\frac{\lambda\gamma
f'}{\sqrt{\alpha\gamma r^4}}\biggr]^2-\frac{1}{\alpha
r^2}-\frac{E}{\alpha J^2}.
\end{equation}
By substituting (\ref{Schwarzschild}) and
$dr/d\phi=-(1/u^2)du/d\phi$ into (\ref{neworbital}), we get after
some manipulation:
\begin{equation}
\label{finalorbital}
\frac{d^2u}{d\phi^2}+u=\frac{EGM}{J^2}-\frac{E\lambda
sG^2M^2}{3r_0^2J^2}\exp\biggl(-\frac{1}{r_0u}\biggr)\biggl(1+\frac{1}{r_0u}\biggr)
+3GMu^2,
\end{equation}
where $r_0=1/\mu$.

For material test particles $E=1$ and we obtain
\begin{equation}
\label{materialorbit}
\frac{d^2u}{d\phi^2}+u=\frac{GM}{J^2}+3GMu^2-\frac{K}{J^2}\exp\biggl(-\frac{1}{r_0u}\biggr)
\biggl(1+\frac{1}{r_0u}\biggr),
\end{equation}
where $K=\lambda sG^2M^2/3r_0^2$. On the other hand, for massless
photons $ds^2=0$ and $E=0$ and (\ref{finalorbital}) gives
\begin{equation}
\label{photons} \frac{d^2u}{d\phi^2}+u=3GMu^2.
\end{equation}

\section{Galaxy Rotational Velocity Curves}

A possible explanation of the galactic rotational velocity curves
problem has been obtained in NGT~\cite{Sokolov}. We shall now
obtain an equivalent explanation of the rotational velocity curves
from the present MSTG theory. From the radial acceleration derived
from (\ref{Yukawa}) experienced by a test particle in a static,
spherically symmetric gravitational field due to a point source,
we obtain
\begin{equation}
\label{accelerationlaw}
a(r)=-\frac{G_{\infty}M}{r^2}+\sigma\frac{\exp(-r/r_0)}{r^2}
\biggl(1+\frac{r}{r_0}\biggr).
\end{equation}
Here, $G_{\infty}$ is defined to be the {\it effective}
gravitational constant at infinity
\begin{equation}
\label{renormG}
G_{\infty}=G_0\biggl(1+\sqrt{\frac{M_0}{M}}\biggr),
\end{equation}
where $G_0$ is Newton's ``bare'' gravitational constant. Moreover,
$M$ is the total mass and $M_0$ is a parameter related to the
strength of the coupling of the skew field to matter.

This conforms with our definition of $G$ in
Eq.(\ref{gravirenorm}), which requires that the effective $G$ be
renormalized in order to guarantee that (\ref{accelerationlaw})
reduces to the Newtonian acceleration
\begin{equation}
\label{Newtonianacceleration}
a_{\rm Newton}=-\frac{G_0M}{r^2}
\end{equation}
at small distances $r\ll r_0$. The constant $\sigma$ is given by
(we reinstate $c$):
\begin{equation}
\label{sigma2}
\sigma=\frac{\lambda s G_0^2M^2}{3c^2r_0^2}.
\end{equation}

The integration constant $s$ in (\ref{sigma2}) is dimensionless
and can be modelled as
\begin{equation}
\label{sparameter} s=gM^b,
\end{equation}
where $g$ is a coupling constant and $b$ is a dimensionless
constant. We choose $b=-3/2$ and set $\lambda
gG_0/3c^2r_0^2=\sqrt{M_0}$. The choice of $b=-3/2$ yields for
galaxy dynamics an approximate Tully-Fisher law~\cite{Tully}. The
expression (\ref{sparameter}) for the parameter $s$ is a
phenomenological model for the coupling of the skew field to
matter, which does not at present have a fundamental
interpretation.

We obtain the acceleration on a point particle
\begin{equation}
\label{accelerationlaw2}
a(r)=-\frac{G_{\infty}M}{r^2}+G_0\sqrt{MM_0}\frac{\exp(-r/r_0)}{r^2}
\biggl(1+\frac{r}{r_0}\biggr).
\end{equation}
By using (\ref{renormG}), we can express the modified acceleration
in the form
\begin{equation}
\label{accelerationlaw3} a(r)=-\frac{G_0M}{r^2}
\biggl\{1+\sqrt\frac{M_0}{M}\biggl[1-\exp(-r/r_0)\biggl(1+\frac{r}{r_0}\biggr)
\biggr]\biggr\}.
\end{equation}

In section 10, we shall explain the running with distance scale of
the effective gravitational constant $G=G(r)$ and the skew field
coupling constant $s=s(r)$ in an RG flow, effective MSTG action
framework in which RG phase space trajectories possess Gaussian
and non-Gaussian fixed points.

We can rewrite (\ref{accelerationlaw3}) in the form
\begin{equation}
\label{runG} a(r)=-\frac{G(r)M}{r^2},
\end{equation}
where
\begin{equation}
\label{runningG}
G(r)=G_0\biggl\{1+\sqrt\frac{M_0}{M}\biggl[1-\exp(-r/r_0)\biggl(1+\frac{r}{r_0}\biggr)
\biggr]\biggr\}.
\end{equation}
Thus, $G(r)$ describes the running with distance of the effective
gravitational constant in the RG flow scenario. Moreover, the
parameter $M_0$, which is a phenomenological description of the
coupling strength of the skew field to matter will also run with
distance in the RG flow scenario.

We apply (\ref{accelerationlaw3}) to explain the flatness of
rotation curves of galaxies, as well as the approximate
Tully-Fisher law~\cite{Tully}, $v^4\propto G_0M\propto L$, where
$v$ is the rotational velocity of a galaxy, $M$ is the galaxy mass
\begin{equation}
\label{Mass}
 M=M_*+M_{HI}+M_{DB}+M_f,
\end{equation}
and $L$ is the galaxy luminosity. Here, $M_*,M_{HI}$, $M_{DB}$ and
$M _f$ denote the visible mass, the mass of neutral hydrogen,
possible dark baryon mass and gas, and the mass from the skew
field energy density, respectively.

The rotational velocity of a star $v$ obtained from
$v^2(r)/r=a(r)$ is given by
\begin{equation}
\label{rotvelocity}
v=\sqrt{\frac{G_0M}{r}}\biggl\{1+\sqrt{\frac{M_0}{M}}\biggl[1-\exp(-r/r_0)
\biggl(1+\frac{r}{r_0}\biggr)\biggr]\biggr\}^{1/2}.
\end{equation}

Let us define the relation between parameters $M_0$ and $r_0$ in
terms of the magnitude of the constant acceleration
\begin{equation}
\label{specialacceleration}
a_0=\frac{G_0M_0}{r^2_0}.
\end{equation}
We assume that for galaxies and clusters of galaxies this
acceleration is determined by
\begin{equation}
\label{Hubbleacceleration} a_0=cH_0.
\end{equation}
Here, $H_0$ is the
current measured Hubble constant $H_0=100\, h\, {\rm km}\,
s^{-1}\,{\rm Mpc}^{-1}$ where $h=(0.71\pm 0.07)$~\cite{pdata}.
This gives
\begin{equation}
\label{speciala} a_0=6.90\times 10^{-8}\,{\rm cm}\, s^{-2}.
\end{equation}
We note that $a_0=cH_0\sim (\sqrt{\Lambda/3})c^2$, so there is an
interesting connection between the parameters $M_0$, $r_0$ and the
cosmological constant $\Lambda$.

A good fit to low surface brightness and high surface brightness
galaxy data is obtained with the parameters
\begin{equation}
\label{parameters} M_0=9.60\times
10^{11}M_{\odot},\quad r_0=13.92\,{\rm kpc}=4.30\times
10^{22}\,{\rm cm}
\end{equation}
and $M$ (or the mass-to-light ratio $M/L$). In (\ref{parameters})
the parameter $r_0$ is not independent of $M_0$. By using
(\ref{speciala}) and substituting $M_0=9.60\times
10^{11}M_{\odot}$ into (\ref{specialacceleration}), we obtain the
$r_0$ in (\ref{parameters}). Thus, we fit the galaxy rotation
curve data with one parameter $M_0$ and the total galaxy mass $M$.
Since we are using an equation of motion for point particle
sources, we are unable to fit the cores of galaxies without
supplementing the acceleration formula with a galaxy core model
based on a mass distribution.

Let us now describe a model of a spherically symmetric galaxy with
a core density of visible matter, $\rho_c(r)$, within a core
radius $r < r_c$. The acceleration law takes the form
\begin{equation}
\label{coremodel}
a(r)=-\frac{G_0{\cal
M}(r)}{r^2}\biggl\{1+\sqrt{\frac{M_0}{M}}\biggl[1-\exp(-r/r_0)\biggl(1+\frac{r}{r_0}
\biggr)\biggr]\biggr\},
\end{equation}
where
\begin{equation}
{\cal M}(r)=4\pi \int_0^rdr'r'^2\rho_c(r')
\end{equation}
is the ordinary matter inside the luminous core of the galaxy
described by a ball of radius $r=r_c$. Inside the core radius
$r_c$ the dynamics is described by Newtonian theory. Moreover, the
acceleration outside the core radius $r
> r_c$ is described by (\ref{accelerationlaw3}), while $M$ is
determined by (\ref{Mass}). A simple model for ${\cal M}(r)$ is
given by\footnote{We can choose to solve a Poisson equation for
the potential that gives a more accurate description of the core
behavior of the galaxy, including bulge effects and other detailed
features of the galaxy associated with a measured luminosity
distribution. However, the model considered here yields a
reasonable description of the velocity curves in the core regions
of the galaxies.}
\begin{equation}
\label{coreacceleration}
 {\cal M}(r)=M\biggl(\frac{r}{r_c+r}\biggr)^\beta,
\end{equation}
where $M$ is the constant total mass and $\beta={\rm constant}$.
The rotational velocity derived from the acceleration law
(\ref{coremodel}) is
\begin{equation}
\label{rotvelocity2}
v=\sqrt{\frac{G_0M}{r}}\biggl(\frac{r}{r_c+r}\biggr)^{\beta/2}
\biggl\{1+\sqrt{\frac{M_0}{M}}\biggl[1-\exp(-r/r_0)
\biggl(1+\frac{r}{r_0}\biggr)\biggr]\biggr\}^{1/2}.
\end{equation}

The modified acceleration law (\ref{coremodel}) can be compared to
the Newtonian law using (\ref{coreacceleration}):
\begin{equation}
\label{coreNewton}
 a_{\rm Newton}(r)=-\frac{G_0{\cal M}(r)}{r^2}.
\end{equation}

The fits to the galaxy rotation curves $v$ in km/s versus the
galaxy radius $r$ in kpc are shown in Fig. 1. The acceleration law
is given for point particles by (\ref{accelerationlaw3}). The data
are obtained from ref.~\cite{McGaugh}.

In Fig. 2, fits to two dwarf galaxies (dSph) are shown. We assume
that the relation between the velocity dispersion $\sigma$ and the
rotational velocity $v$ takes the simple form in e.g. an
isothermal sphere model for which $v\sim \sqrt{2}\sigma$. The
error bars on the data~\cite{Lokas} for the velocity dispersions
are large, and in the case of Draco, due to the small radial range
$0.1\,{\rm kpc} < r < 0.6\,{\rm kpc}$, the Newtonian curve for
\begin{equation}
v=\sqrt{\frac{G_0M}{r}},
\end{equation}
cannot be distinguished within the errors from the MSTG
prediction. However, it is noted that the MSTG prediction for $v$
appears to flatten out as $r$ increases. For Draco
$M/L=28.93+50.30(9.58)(M_{\odot}/L_{\odot})$, whereas for Fornax
$M/L=1.79+0.72(0.40)(M_{\odot}/L_{\odot})$. There is also an
expected large error in the distance estimates to the dSph.
Another serious potential source of error is that it is assumed
that dSph galaxies are in dynamical equilibrium. The two studied
here are members of the Local Group and exist in the gravitational
field of a larger galaxy, the Milky Way. Thus, the tidal
interactions with the larger galaxy are expected to affect the
dynamics of dSph galaxies and the interpretations of velocity
dispersions~\cite{Mateo}. These issues and others for dSph
galaxies are critically considered in the context of dark matter
models by Kormendy and Freeman~\cite{Kormendy}.

We have also included a fit to the elliptical galaxy NGC 3379 as
shown in Fig. 3. The elliptical galaxy NGC 3379 has been the
source of controversy recently~\cite{Romanowsky}. The velocities
of elliptical galaxies are randomly distributed in the galaxy.
However, the gravitational potential that would be experienced by
a test particle star or planetary nebula in circular rotation
about the center of the galaxy can be extracted from the
line-of-sight velocity dispersion profiles. The data for $R/R_{\rm
eff}> 0.5$ refer to planetary nebula.

We use the mean values of the extracted rotational velocities for
NGC 3379, obtained by Romanowsky et al.~\cite{Romanowsky} and find
that our predicted rotational velocities agree well with their
data as shown in Fig. 3. According to Romanowsky et al. there
appears to be a dearth of dark matter in the elliptical galaxy
which needs to be explained by dark matter models and N-body
cosmological simulations. For Milgrom's
MOND~\cite{Milgrom,Bekenstein,Aguirre} it is argued by Sanders and
Milgrom~\cite{Sanders} that NGC 3379 is marginally outside the
MOND regime with an acceleration $a\sim (a_0)_{\rm Milgrom}\sim
1.2\times 10^{-8}\,{\rm cm}\,{\rm s}^{-2}$, so the MOND prediction
for elliptical galaxies should be closer to the Newtonian-Kepler
prediction for the rotational velocity curve. Romanowsky et al.
also give data for the two elliptical galaxies NGC 821 and NGC
4494, but the intrinsic circular velocities inferred from the
line-of-sight velocity dispersion profile data are not given by
the authors, although the trends of the data are similar to NGC
3379.

A fit to the data for the globular cluster $\omega$ Centauri is
shown in Fig. 3. The data is from McLaughlin and
Meylan~\cite{Meylan}. We use the velocity dispersion data and
assume that the data is close to the rotational velocity curves
associated with the velocity dispersion $\sigma_p$ i.e. the
isothermal sphere model relation $v\sim \sqrt{2}\sigma_p$ holds.
The fit to the data reveals that the predicted rotational velocity
cannot be distinguished from the Newtonian-Kepler circular
velocity curve within the orbital radius of the data. The authors
conclude that there appears to be no room for dark matter, whereas
the our results agree well with the data. For Milgrom's MOND, the
magnitude of the acceleration is larger than the MOND upper limit
$(a_0)\sim 10^{-8}\,{\rm cm}\,/{\rm sec}$ and we expect to obtain
a Newtonian-Kepler rotation curve.

In Fig. 4, we show fits to the four galaxies NGC 1560, NGC 2903,
NGC 4565 and NGC 5055 using the modified acceleration law
(\ref{coremodel}) including the galaxy core with $\beta=2$. As can
be seen from the data fitting, the agreement with the core and
extended rotation curves data ~\cite{Begeman,Sofue} is good for
the two fitting parameters $M$ and the core radius $r_c$.

In Fig. 5, we display a 3-dimensional plot of $v$ versus the range
of distance $0.1\,{\rm kpc} < r < 10\,{\rm kpc}$ and the range of
total galaxy mass $M$ used in the fitting of rotational velocity
data. The red surface shows the Newtonian values of the rotational
velocity $v$, while the dark surface displays the prediction for
$v$ obtained from our gravity theory.

Table 1 displays the values of the total mass $M$ used to fit the
galaxies and the mass-to-light ratios $M/L$ estimated from the
data in references given in~\cite{McGaugh}, using the non-core
acceleration model (\ref{accelerationlaw3}).

In Milgrom's phenomenological model~\cite{Milgrom,Bekenstein} we
have
\begin{equation}
\label{Milgromv}
v^4=G_0M(a_0)_{\rm Milgrom},
\end{equation}
where $(a_0)_{\rm Milgrom}=1.2\times 10^{-8}\,{\rm cm}\, s^{-2}$.
We see that (\ref{Milgromv}) predicts that the rotational velocity
is constant out to an infinite range and the rotational velocity
does not depend on a distance scale, but on the magnitude of the
acceleration $(a_0)_{\rm Milgrom}$\footnote{In formulations of
Milgrom's model, the modified Newtonian potential has a
logarithmic dependence on $r$}. In contrast, our modified
acceleration formula does depend on the radius $r$ and the
distance scale $r_0$ which for galaxies is fixed by the formula
(\ref{Hubbleacceleration}). The MSTG velocity curve asymptotically
becomes the same as the Newtonian-Kepler prediction as
$r\rightarrow\infty$:
\begin{equation}
\label{asympvelocity} v\sim \sqrt{G_{\infty}M/r},
\end{equation}
where $G_{\infty}$ is the renormalized value of Newton's constant.

Using the Sloan Digital Sky Survey (SDSS), Prada et
al.~\cite{Prada} have studied the velocities of satellites
orbiting isolated galaxies. They detected approximately 3000
satellites, and they found that the line-of-sight velocity
dispersion of satellites declines with distance to the primary.
The velocity was observed to decline to a distance of $\sim 350$
kpc for the available data. This result contradicts the constant
velocity prediction (\ref{Milgromv}) of MOND, but agrees with the
MSTG prediction (\ref{asympvelocity}). It also agrees with the
cosmological models which predict mass profiles of dark matter
halos at large distances. During the last two decades of numerical
modelling of galaxy formation, they have produced a density
profile of dark matter halos, $\rho\propto 1/r^3$ at large radii,
which does not depend on the nature of the dark
matter~\cite{Avila}.

Although it appears that the MOND prediction for the velocity
distribution of satellite galaxies disagrees with the Prada et al.
data, it can be argued that if a system is embedded in an external
gravitational field of Newtonian acceleration $a_{\rm ext}$, then
if the internal Newtonian acceleration of the system $a_{\rm
Newton}$ satisfies $a_{\rm Newton} < a_{\rm ext} < a_0$, then the
modified Milgrom acceleration is $a_{\rm Newton}(a_0/a_{\rm ext})$
instead of $\sqrt{a_0a_{\rm Newton}}$. The acceleration becomes
Kepler-Newtonian and anisotropic with an increased effective
gravitational constant~\cite{Aguirre}. However, the MSTG
prediction of the velocity dispersion does not depend sensitively
on knowing the modelling of an external gravitational field
influence on the satellite galaxies.

\section{Local and Solar System Observations}

We obtain from Eq.(\ref{materialorbit}) the orbit equation (we
reinsert the speed of light c):
\begin{equation}
\label{particleorbit}
\frac{d^2u}{d\phi^2}+u=\frac{GM}{c^2
J^2}-\frac{K}{J^2}\exp(-r/r_0)\biggl[1
+\biggl(\frac{r}{r_0}\biggr)\biggr]+\frac{3GM}{c^2}u^2,
\end{equation}
where now $K=\lambda sG^2M^2/3c^4r_0^2$.
Using the large $r$ weak field approximation, and the expansion
\begin{equation}
\exp(-r/r_0)=
1-\frac{r}{r_0}+\frac{1}{2}\biggl(\frac{r}{r_0}\biggr)^2+...
\end{equation}
we obtain the orbit equation for $r\ll r_0$:
\begin{equation}
\label{orbitperihelion}
\frac{d^2u}{d\phi^2}+u=N+3\frac{GM}{c^2}u^2,
\end{equation}
where
\begin{equation}
N=\frac{GM}{c^2J_N^2}-\frac{K}{J_N^2}.
\end{equation}

We can solve Eq.(\ref{orbitperihelion}) by perturbation theory and
find for the perihelion advance of a planetary orbit
\begin{equation}
\label{perihelionformula} \Delta\omega=\frac{6\pi}{c^2L}
(GM_{\odot}-c^2K_{\odot}),
\end{equation}
where $J_N=(GM_{\odot}L/c^2)^{1/2}$, $L=a(1-e^2)$ and $a$ and $e$
denote the semimajor axis and the eccentricity of the planetary
orbit, respectively.

We now use the running of the effective gravitational coupling
constant $G=G(r)$, determined by (\ref{runningG}) and find that
for the solar system $r\ll r_0=14\, {\rm kpc}$, we have $G\sim
G_0$ within the experimental errors for the measurement of
Newton's constant $G_0$. We choose for the solar system
$K_{\odot}\ll 1.5\,{\rm km}$ and use $G=G_0$ to obtain from
(\ref{perihelionformula}) a perihelion advance of Mercury in
agreement with GR.

For terrestrial experiments and orbits of satellites, we have also
that $G\sim G_0$ for $r\sim r_{\oplus}$ where $r_{\oplus}\ll
r_0=14\,{\rm kpc}$ is the radius of Earth. We then achieve
agreement with all gravitational terrestrial experiments including
E\"otv\"os free-fall experiments and ``fifth force'' experiments.

For the binary pulsar PSR 1913+16 the formula
(\ref{perihelionformula}) can be adapted to the periastron shift
of a binary system. Combining this with the MSTG gravitational
wave radiation formula, which will approximate closely the GR
formula, we can obtain agreement with the observations for the
binary pulsar.  The mean orbital radius for the binary pulsar is
equal to the projected semi-major axis of the binary, $\langle
r\rangle_N=7\times 10^{10}\,{\rm cm}$, so that $\langle
r\rangle_N\ll 14\, {\rm kpc}$. Thus, $G=G_0$ within the
experimental errors and agreement with the binary pulsar data for
the periastron shift is obtained for $K_N\ll 4.2\,{\rm km}$.

For a massless photon $E=0$ and we have
\begin{equation}
\label{lightbending}
\frac{d^2u}{d\phi^2}+u=3\frac{GM}{c^2}u^2.
\end{equation}
For the solar system using $G=G_0$ within the experimental errors
gives the light deflection:
\begin{equation}
\Delta_{\odot}=\frac{4G_0M_{\odot}}{c^2R_{\odot}}
\end{equation}
in agreement with GR.

\section{Galaxy Clusters and Lensing}

We can assess the existence of dark matter of galaxies and
clusters of galaxies in two independent ways: from the dynamical
behavior of test particles through the study of extended rotation
curves of galaxies, and from the deflection and focusing of
electromagnetic radiation, e.g., gravitational lensing of clusters
of galaxies. The light deflection by gravitational fields is a
relativistic effect, so the second approach provides a way to test
the relativistic effects of gravitation at the extra-galactic
level. It has been shown that for conformal tensor-scalar gravity
theories the bending of light is either the same or can even be
weaker than predicted by GR~\cite{Bekenstein,Bekenstein2}. To
remedy this problem, Bekenstein~\cite{Bekenstein} has recently
formulated a relativistic description of Milgrom's MOND model,
including a time-like vector field as well as two scalar fields
within a GR metric scenario. However, the time-like vector field
violates local Lorentz invariance and requires preferred frames of
reference.

The bending angle of a light ray as it passes near a massive
system along an approximately straight path is given to lowest
order in $v^2/c^2$ by
\begin{equation}
\label{lensingformula} \theta=\frac{2}{c^2}\int\vert
a^{\perp}\vert dz,
\end{equation}
where $\perp$ denotes the perpendicular component to the ray's
direction, and dz is the element of length along the ray and $a$
denotes the acceleration. The best evidence in favor of dark
matter lensing is the observed luminous arcs seen in the central
regions of rich galaxy clusters~\cite{Blanford}. The cluster
velocity dispersion predicted by the observed arcs is consistent
within errors with the observed velocity dispersion of the cluster
galaxies. This points to a consistency between the virial mass and
the lensing mass, which favors the existence of dark matter.

From (\ref{lightbending}), we obtain the
light deflection
\begin{equation}
\Delta=\frac{4GM}{c^2R}=\frac{4G_0{\overline M}}{c^2R},
\end{equation}
where
\begin{equation}
{\overline M}=M\biggl(1+\sqrt{\frac{M_0}{M}}\biggr).
\end{equation}
The value of ${\overline M}$ follows from (\ref{runningG}) for
clusters as $r\gg r_0$ and
\begin{equation}
G(r)\rightarrow
G_{\infty}=G_0\biggl(1+\sqrt{\frac{M_0}{M}}\biggr).
\end{equation}
We choose for a cluster $M_0=3.6\times 10^{15}\,M_{\odot}$ and a
cluster mass $M_{\rm cl}\sim 10^{14}\,M_{\odot}$, and obtain
\begin{equation}
\biggl(\sqrt{\frac{M_0}{M}}\biggr)_{\rm cl}\sim 6.
\end{equation}
We see that ${\overline M}\sim 7M$ and we can explain the increase
in the light bending without exotic dark matter.

From the formula Eq.(\ref{accelerationlaw3}) for $r\gg r_0\sim
14\,{\rm kpc}$ we get
\begin{equation}
a(r)=-\frac{G_0\overline M}{r^2}.
\end{equation}
We expect to obtain from this result a satisfactory description
of lensing phenomena using Eq.(\ref{lensingformula}).

The scaling with distance of the renormalized gravitational
constant is seen to play an important role in describing
consistently the solar system and the galaxy and cluster dynamics,
without the postulate of exotic dark matter. In the next section,
we outline a framework for an effective, running Newtonian
gravitational constant $G=G(k)$ and the skew field coupling
constant $\gamma_c(k)$, based on a renormalization group approach
to gravity theory.

\section{Renormalization Group Flow Gravity Theory and Running Constants}

We have postulated a renormalized Newtonian gravitational constant
$G=G_0Z$ in our action Eq.(\ref{gravaction}), and in the
phenomenology for galaxy dynamics
\begin{equation}
Z=1+\sqrt{\frac{M_0}{M}}.
\end{equation}
We shall now implement the running of the effective $G$ and the
$F_{\mu\nu\lambda}$ field coupling constant $\gamma_c$ by using
renormalization flow arguments in which the ``classical'' coupling
constants $G$ and $\gamma_c$ possess a scale dependent running
behavior obtained by solving appropriate RG equations. This
approach to quantum gravity and its astrophysical implications for
the infrared (IR) low energy or large scale dependence of gravity
has been applied to Einstein
gravity~\cite{Reuter,Reuter2,Reuter3,Shapiro} with the intent to
explain galaxy rotation curves without dominant dark matter.

We shall postulate scale dependent effective actions
$\Gamma_k[g_{\mu\nu}]$ and $\Gamma_k[A_{\mu\nu}]$, corresponding
to ``coarse-grained'' free energy functionals, which define an
effective field theory valid at the mass scale $k$ or length scale
$\ell=1/k$. Then the $\Gamma_k$ are the bare actions obtained by
integrating out all quantum fluctuations with momenta larger than
an infrared cutoff $k_{IR}$, or wavelengths smaller than $\ell$.
The $\Gamma_k$ are determined by functional differential equations
for the RG flow, including the dimensionless coupling constants
${\bar G}(k)=k^2G(k), \gamma_c(k)$, the cosmological constant
$\lambda(k)=\Lambda(k)/k^2$ and the mass ${\bar\mu}=\mu(k)/k$. The
dimensionless constant $G(k)\Lambda(k)={\bar G}(k)\lambda(k)$ is
the quantity that is determined by measurement. The flow equations
are described by a system of ordinary coupled differential
equations:
\begin{equation}
k\frac{d}{dk}{\bar G}(k)=\beta_{\bar G}({\bar G},{\gamma}_c,
\lambda,{\bar\mu}),\quad
k\frac{d}{dk}{\gamma}_c(k)=\beta_{{\gamma}_c}({\bar
G},{\gamma}_c,\lambda,{\bar\mu}),
$$ $$
k\frac{d}{dk}\lambda(k)=\beta_\lambda({\bar
G},\gamma_c,\lambda,{\bar\mu}),\quad
k\frac{d}{dk}\mu(k)=\beta_{\bar\mu}({\bar
G},\gamma_c,\lambda,{\bar\mu}).
\end{equation}
To solve the problem, we are required to restrict the RG flow to a
finite-dimensional subspace~\cite{Reuter}, corresponding to a
truncated theory space. In the truncated space only the coupling
constants ${\bar G}$ and ${\gamma}_c$, the cosmological constant
$\lambda$ and the mass $\bar\mu$ are considered.

It is assumed that there exists an RG trajectory $(G(k)$,\break
$\gamma_c(k),\Lambda(k),\mu(k))$ and that this gives rise to
running coupling constants $G(k), \gamma_c(k)$, a running
cosmological constant $\Lambda(k)$ and mass $\mu(k)$ with a cutoff
$k=k(x)$. The latter converts the scale dependences of $G(k),
\gamma_c(k)$, $\Lambda(k)$ and $\mu(k)$ to a spacetime position
dependence~\cite{Reuter,Reuter2,Reuter3}:
\begin{equation}
G(x)\equiv G(k=k(x)),\quad \gamma_c(x)\equiv
\gamma_c(k=k(x)),\quad \Lambda\equiv \Lambda(k=k(x)),
$$ $$
\mu\equiv\mu(k=k(x)).
\end{equation}

We shall follow the procedure of Reuter and
Weyer~\cite{Reuter,Reuter2,Reuter3} and rewrite the action $S_G$
in (\ref{gravaction}) as
\begin{equation}
\label{action2} {\bar S}_G=\frac{1}{16\pi G(x)}\int
d^4x\sqrt{-g}[R-2\Lambda(x)],
\end{equation}
and the skew field action as
\begin{equation}
\label{skewaction2} {\bar S}_F=\int
d^4x\sqrt{-g}(\frac{1}{12}F_{\mu\nu\rho}F^{\mu\nu\rho}
-\frac{1}{4}\mu^2(x)A_{\mu\nu}A^{\mu\nu}).
\end{equation}

We have required the skew field mass $\mu=\mu(x)$ to be a function
of the spacetime coordinates, allowing for a mass renormalization.
The matter action $S_M$ is supplemented by an energy-momentum
tensor $\Theta_{\mu\nu}$:
\begin{equation}
\frac{1}{\sqrt{-g}}\biggl(\frac{\delta S_\Theta}{\delta
g^{\mu\nu}}\biggr)=-\frac{1}{2}\Theta_{\mu\nu},
\end{equation}
which describes the energy-momentum associated with the $G(x)$,
$\gamma_c(x)$, $\Lambda(x)$ and $\mu(x)$. The new field equations
that replace (\ref{Gequation}) are
\begin{equation}
\label{Gequation2} G_{\mu\nu}+\Lambda g_{\mu\nu}=8\pi G{\bar
T}_{\mu\nu}.
\end{equation}
The modified energy-momentum tensor ${\bar T}^{\mu\nu}$, which
includes the contribution from $\Theta^{\mu\nu}$, must satisfy the
generalized Bianchi identities, which follow from the
diffeomorphism of the action:
\begin{equation}
\nabla_{\nu}{\bar T}^{\mu\nu}=0.
\end{equation}
These conservation laws lead to constraints on the
$\Theta^{\mu\nu}$. The fields $G(x)$, $\gamma_c(x)$, $\Lambda(x)$
and $\mu(x)$ are not varied in the effective actions ${\bar S}_G$
and ${\bar S}_F$ as they are treated as ``background external
fields'' and they do not have corresponding field
equations~\cite{Reuter,Reuter2,Reuter3}. Thus, the actions do not
contain kinetic energy terms corresponding to the fields $G(x),
\gamma_c(x)$, $\Lambda(x)$ and $\mu(x)$.

The flow equations are described by the running constants
\begin{equation}
G(k)={\bar G}(k)/k^2,\quad \gamma_c(k),\quad \Lambda(k)=\lambda
k^2,\quad \mu(k)={\bar\mu}(k)k.
\end{equation}
Every solution $({\bar G}(k),\gamma_c(k),\lambda(k),{\bar\mu(k)})$
of the truncated flow equations is associated with the
one-parameter family of action functionals
\begin{equation}
\Gamma_k[g_{\mu\nu}]=\frac{1}{16\pi G(k)}\int
d^4x\sqrt{-g}[R-2\Lambda(k)],
\end{equation}
\begin{equation}
\Gamma_k[A_{\mu\nu}]=\int
d^4x\sqrt{-g}(\frac{1}{12}F_{\mu\nu\rho}F^{\mu\nu\rho}
-\frac{1}{4}\mu^2(k)A_{\mu\nu}A^{\mu\nu}).
\end{equation}

The effective average action formalism follows the RG flow from
the bare action $\Gamma_{k=\infty}=S$, corresponding to the
initial condition for the $\Gamma_k$-flow equation, down to $k=0$
and $\Gamma_{k=0}=\Gamma$ corresponding to the effective action.
We have to solve the effective equations of motion
\begin{equation}
\frac{\delta\Gamma_k[g_{\mu\nu}]}{\delta g_{\mu\nu}}=0,
\end{equation}
\begin{equation}
\frac{\delta\Gamma_k[A_{\mu\nu}]}{\delta A_{\mu\nu}}=0.
\end{equation}

In practice, we truncate the space of solutions and for the UV
sector, simple local truncations are sufficient to describe the
physical system, but in the IR limit $k\rightarrow 0$ {\it
nonlocal} terms must be included in the truncation procedure.

We shall presently just consider the running of $G(k)$ and the
coupling constant $\gamma_c(k)$. The RG flow equations are
dominated by two fixed points ${\bar G}_*$ and
${\bar\gamma}_{c*}$: Gaussian fixed points at ${\bar
G}_*={\gamma}_{c*}=0$ and non-Gaussian ones with ${\bar G}_*
> 0$ and ${\gamma}_{c*} > 0$. The high-energy, short distance
behavior of the quantum gravity theory is governed by the
non-Gaussian fixed points, so that for $k\rightarrow\infty$ all
the RG trajectories run into these fixed points and lead to a
quantum gravity theory by taking the ultra-violet (UV) cutoff
along a trajectory running into the fixed points. This would lead
to a non-perturbatively renormalizable quantum gravity
theory~\cite{Weinberg2,Lauscher,Souma,Perini,Litim}. The
conjectured existence of such a non-perturbatively renormalizable
gravity theory still has to be demonstrated.

Both the couplings $G(k)$ and $\gamma_c(k)$ are asymptotically
free coupling constants as in quantum chromodynamics (QCD), so
that they vanish for $k\rightarrow\infty$. We see that if we
interpret $\ell=1/k$ as a distance scale, then both $G(k)$ and
$\gamma_c(k)$ increase all the way from the UV energy scale to the
infrared (IR) energy scale as $k\rightarrow 0$. This corresponds
to a quantum gravity {\it anti-screening}, playing the analogous
role to the anti-screening in Yang-Mills QCD. The RG trajectory
ending at the fixed point ${\bar G}(k=0)=0$ can be described by
the scaling law
\begin{equation}
G(k)=G_0\biggl(1+\frac{g_{IR}}{k^2}\biggr).
\end{equation}
This interpolates between the constant Newtonian value $G(k_{\rm
lab})=G_0$ at large $k$ corresponding to short distances and the
increasingly larger values of G(k) at larger distances for small
$k$. The difference between $G$ at $k=0$ and at a typical
laboratory scale, $k\sim 1/({\rm meter})$ is negligible. Any UV
renormalization effects are negligible for laboratory, galaxy and
cosmological scales.

The gravitational coupling constant will display a running power
law behavior:
\begin{equation}
G(k)=\frac{{\bar G}_*}{k^2}.
\end{equation}
For the spherically symmetric static solution, the running of
$G(r)$ is described by Eq.(\ref{runningG}), so that as r increases
to galactic distances $G(r)$ increases until it reaches the
constant value
\begin{equation}
G_{\infty}=G_0\biggl[1+\sqrt{\frac{M_0}{M}}\biggr].
\end{equation}

A more detailed analysis of RG flow equations in terms of our
effective actions will be presented elsewhere.

We note that if we had considered an effective RG framework based
on the Einstein-Hilbert action (\ref{gravaction}) without
including a coupling to the skew field $F_{\mu\nu\lambda}$, then
the weak gravitational field acceleration law would take the
effective ``Newtonian form'' (\ref{runG}). Then, the RG flow
effective action formalism is required to predict that $G(r)$
should run in such a way that it produces flat rotation curves for
galaxies that fit all the galaxy data~\cite{Reuter2}. Whether this
is possible is a matter of conjecture. In contrast, the running of
the effective constant $G=G(r)$ in MSTG theory is determined to a
large extent by the field equations of the theory for weak fields
and leads to good agreement with the available galaxy rotation
curve data.

\section{Cosmology Without Cold Dark Matter}

We shall assume that the universe is isotropic and homogeneous at
large scales and adopt the FLRW line element
\begin{equation}
ds^2=dt^2-R^2(t)\biggl[\frac{dr^2}{1-kr^2}+r^2(d\theta^2+\sin^2\theta
d\phi^2)\biggr].
\end{equation}
The Friedmann equations take the form
\begin{equation}
\label{Friedmann} H^2(t)+\frac{k}{R^2(t)}=\frac{8\pi
G\rho_M(t)}{3}+\frac{\Lambda}{3},
\end{equation}
\begin{equation}
\label{Rdoubledot} {\ddot R}(t)=-\frac{4\pi
G}{3}[\rho_M(t)+3p_M(t)]R(t)+\frac{\Lambda}{3}R(t),
\end{equation}
where $H(t)={\dot R(t)}/R(t)$.

For a homogeneous and isotropic universe the skew symmetric fields
$A_{\mu\nu}$ and $F_{\mu\nu\lambda}$ are zero when averaged over
large distance scales, since there can be no preferred direction
in the maximally symmetric FLRW spacetime. In Eqs.
(\ref{Friedmann}) and (\ref{Rdoubledot}), we have assumed that
$G=G_0Z$ and the renormalized value of $G$ is obtained from the
running of the effective coupling $G=G(t)$ with time. We can write
(\ref{Friedmann}) for a spatially flat universe $k=0$ in the form:
\begin{equation}
\Omega_M(t)+\Omega_{\Lambda}(t)=1,
\end{equation}
where
\begin{equation}
\Omega_M(t)=\frac{8\pi G\rho_M(t)}{3H^2(t)}\quad
\Omega_{\Lambda}=\frac{\Lambda}{3H^2(t)}.
\end{equation}
We shall impose physical restrictions on the form of the running
of $G(t)$. We observe that from calculations of the production of
helium and deuterium abundances at the time of BBN~\cite{Burles},
it can be shown that $\Omega_B\sim 0.02-0.04$. Moreover, from the
WMAP data the fractions of baryon and photon densities at the
surface of last scattering show that $\Omega_B\sim
0.04$~\cite{Spergel}. Therefore, we must impose the conditions
that at the time of nucleosynthesis at $t_{BBN}\sim 10^{-7}\,{\rm
yrs}$ (at red shift $z\sim 10^9$), $G(t_{BBN})\sim G_0$ and at the
surface of last scattering (decoupling) at $t_{SLS}\sim 10^5\,{\rm
yrs}$ ($z\sim 10^3$), we require that $G(t_{SLS})\sim G_0$.
Moreover, since the time of the surface of last scattering we
additionally require that $G(t)$ grows until the present time to
the value $G(t_{\rm now})\sim 6G_0$. Given this running of $G$, it
follows that for $t
> t_{SLS}$:
\begin{equation}
\label{matter} \Omega_M\sim 6\Omega_B\sim 0.24,
\end{equation}
and we need only assume a dominant baryon density. Given this
scenario, we do not require a dominant cold dark matter in the
matter dominated era. There can be small contributions to
$\Omega_M$ due to massive neutrinos.

A bound on the possible value of a changing $G$ with time is
obtained from recent spacecraft measurements~\cite{Williams}:
\begin{equation}
\label{dotGbound} \frac{\dot G}{G} =(4\pm 9)\times 10^{-13}\,{\rm
yr}^{-1}.
\end{equation}
In order to apply this bound to determine the amount of variation
of $G(t)$, we need to know the functional form of $G(t)$ as it
runs with time. This must be determined by solving the RG flow
equations for our cosmological model. If the running of $G$ with
time is such that ${\dot G}/G\sim 0$ for a flat behavior of
$G(t)\sim G_0$ up till the surface of last scattering, followed
after the surface of last scattering by a rapid rise to $G(t)\sim
6G_0$ with a subsequent value ${\dot G}/G\sim 0$, then the bound
(\ref{dotGbound}) will not rule out the required renormalized
value $G(t_{\rm now})\sim 6G_0$. In Fig. 6, we display a
conjectured effective $G(t)$ as a function of $t$ in years that is
consistent with (\ref{dotGbound}) and a baryon dominated universe
without exotic cold dark matter.

The value of $G(t_{\rm now})\sim 6G_0$ required to obtain results
consistent with the WMAP data~\cite{Spergel} corresponds to
cosmological distance scales. For the solar system with a distance
scale of order 1-40 AU, the local value of $G(t_{\rm now})\sim
G_0$ to within the experimental errors on the measurements of
$G_0$.

If we take into account the running of the cosmological constant
$\Lambda(k)$ in the RG quantum gravity flow framework, then we can
have $\Lambda(t)$ run with $t$ in (\ref{Friedmann}) and
(\ref{Rdoubledot}) and this would correspond to a quintessence
scenario~\cite{Steinhardt}. It is possible that the RG flow
trajectories that lead to a large distance classical scenario can
solve the cosmological constant problem, for these trajectories
imply a small cosmological constant~\cite{Reuter2,Shapiro}.

A comparison of our baryon dominated matter era cosmological model
with the acoustical peak data obtained from the WMAP
observations~\cite{Spergel} has to be performed to see whether a
satisfactory fit to the data can be obtained. The concordance
$\Lambda CDM$ model agrees well with the data and the present
model must produce fits to the CMB data that succeeds as well.
Moreover, an investigation of the growth of large scale galaxies
and clusters from an initially smooth background of fluctuations
must also be performed. Since the effective $G$ in the
cosmological model is required to increase after the surface of
last scattering to a value $G(t_{\rm now})\sim 6G_0$, then we can
expect that the gravitational potential well will become deeper
and allow for clumping of the baryon matter to form galaxies
without cold dark matter. The results of this investigation will
be reported elsewhere.

\section{Conclusions}

We have developed a gravity theory consisting of a
metric-skew-tensor action that leads to a modified Newtonian
acceleration law that is fitted to galaxy rotation curves. There
is a large enough sample of galaxy data which fits our predicted
MSTG acceleration law to warrant taking seriously the proposal
that the gravity theory can explain the flat rotational velocity
curves of galaxies without exotic dark matter. It is interesting
to note that we can fit the rotational velocity data of galaxies
in the distance range $0.02\,{\rm kpc} < r < 70\,{\rm kpc}$ and in
the mass range $10^5\, M_{\odot}< M < 10^{11}\,M_{\odot}$ without
exotic dark matter halos~\cite{Brownstein}. We have also provided
a model for the behavior of galaxy rotation curves inside the
luminous core of galaxies that predicts well the observed flatness
of the rotation curves as the distance from the core increases.
The lensing of clusters can also be explained by the theory
without exotic dark matter in cluster halos.

Our RG flow effective action description of MSTG quantum gravity
allows for a running of the effective $G$ with distance. The RG
flow framework for the theory is characterized by special RG
trajectories. On the RG trajectory, we identify a regime of
distance scales where solar system gravitational physics is well
described by GR, which is contained in MSTG as an approximate
solution to the field equations. We are able to obtain agreement
with the observations in the solar system, terrestrial
gravitational experiments and the binary pulsar PSR 1913+16.
Strong infrared renormalization effects become visible at the
scale of galaxies and the modified Newtonian potential replaces
exotic dark matter as an explanation of flat rotation curves.
Thus, gravity becomes a ``confining force'' that has significant
predictions for astrophysics and cosmology.

We have demonstrated that the RG flow running of $G$ and MSTG
cosmology can lead to a description of the universe that does not
require dominant, exotic dark matter. Dark energy is described by
an effective time dependent cosmological constant. A detailed
investigation of the MSTG cosmological scenario must be performed
to establish that it can describe the large scale structure of the
universe, account for galaxy formation and big bang
nucleosynthesis and be consistent with the WMAP data. The RG flow
trajectories in MSTG in the IR large classical limit can lead to a
possible solution to the cosmological constant problem, for they
imply a small cosmological constant~\cite{Reuter,Shapiro}.

The skew fields $F_{\mu\nu\sigma}$ have a natural interpretation
in terms of string theory~\cite{Ramond}, so that a possible string
theory interpretation of the astrophysical predictions can be
investigated.

An interesting approach is to solve the MSTG field equations for
gravitational collapse to see whether there are any stable massive
skew field solutions that correspond to conventional Schwarzschild
black holes with event horizons. If the conjectured asymptotically
safe renormalizability holds for MSTG quantum gravity, then we
might expect that $G(s)$ and $\gamma_c(s)$ vanish as the distance
$s=({x\cdot x})^{1/2}\rightarrow 0$, which could lead to a
gravitationally-free Minkowski spacetime without singularities at
$s=0$.

\vskip 0.2 true in {\bf Acknowledgments} \vskip 0.2 true in

This work was supported by the Natural Sciences and Engineering
Research Council of Canada. I thank Hilary Carteret for help with
the use of Maple 9 software and Joel Brownstein, Gilles
Esposito-Far\'ese, Laurent Freidel, Arthur Lue, Gary Mamon, Stacey
McGough, Martin Green, Martin Reuter and Lee Smolin for helpful
discussions.

  \pagebreak \begin{center} Table 1. Values
of the total galaxy mass $M$ used to fit rotational velocity data.
Also shown are the mass-to-light-ratios $M/L$ with $L$ obtained
from ref.~\cite{McGaugh}. \vskip 0.3 true in
\begin{tabular}{|l||c||c|}\hline Galaxy & M($\times
10^{10}M_{\odot}$)& M/L($M_{\odot}/L_{\odot}$)\\ \hline NGC 5533 &
24.2 & 4.29\\ \hline NGC 5907 & 11.8 & 4.92\\ \hline NGC 6503 &
1.36 & 2.84\\ \hline NGC 3198 & 3.0 & 3.33\\ \hline NGC 2403 &
2.37 & 3.0\\\hline NGC 4138 & 2.94 & 3.59\\ \hline NGC 3379 & 5.78 & --\\ \hline M33 & 0.93 & 1.98\\
\hline UGC 6917 & 0.96 & 2.53\\ \hline UGC 6923 & 0.388 & 1.76\\
\hline UGC 6930 & 1.04  & 2.08\\ \hline FORNAX & 0.0026 & 1.86\\ \hline DRACO & 0.00050 & 27.94\\
\hline $\omega$ Centauri & $3.05\times 10^{-5}$ & -- \\\hline
\end{tabular}
\end{center}

\pagebreak

\vskip 0.1 in

\begin{center}
\includegraphics[width=2.5in,height=2.5in]{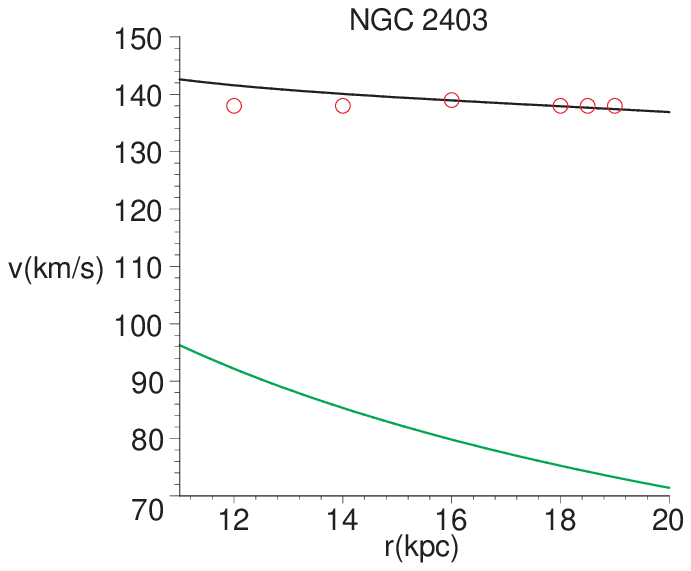}
\includegraphics[width=2.5in,height=2.5in]{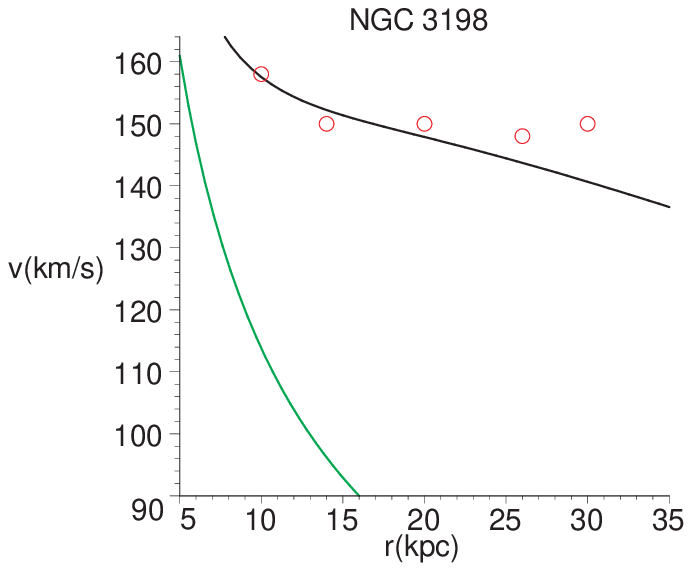}
\end{center}
\vskip 0.1 true in
\begin{center}
\includegraphics[width=2.5in,height=2.5in]{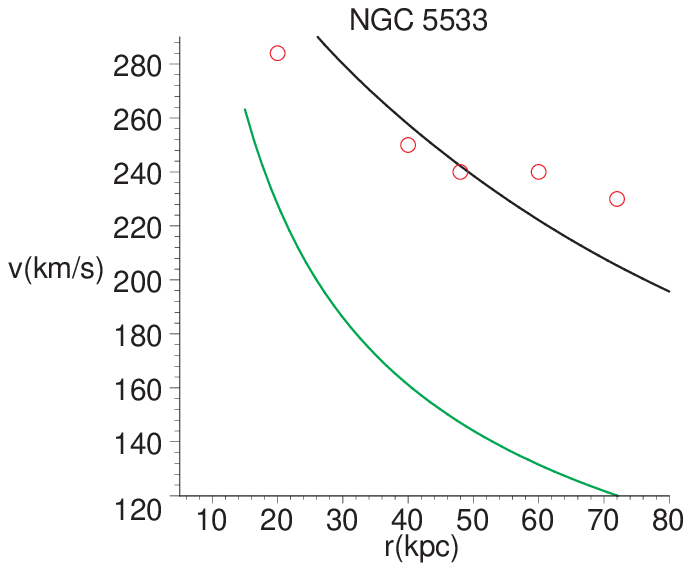}
\includegraphics[width=2.5in,height=2.5in]{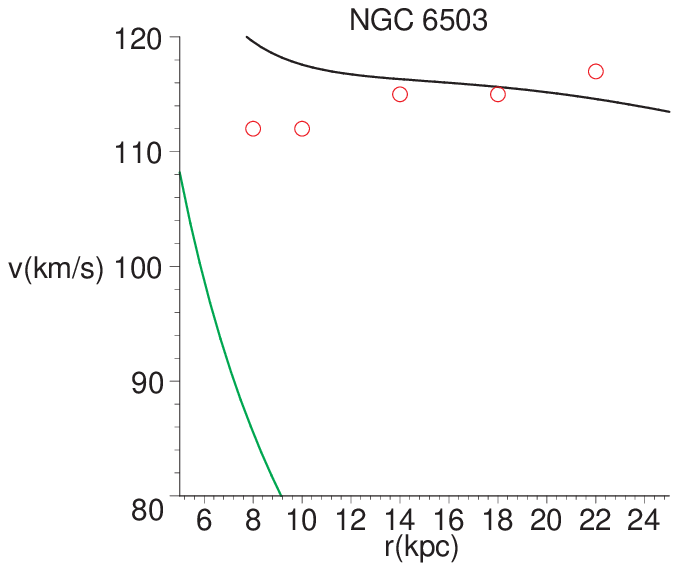}
\end{center}
\vskip 0.1 true in
\begin{center}
\includegraphics[width=2.5in,height=2.5in]{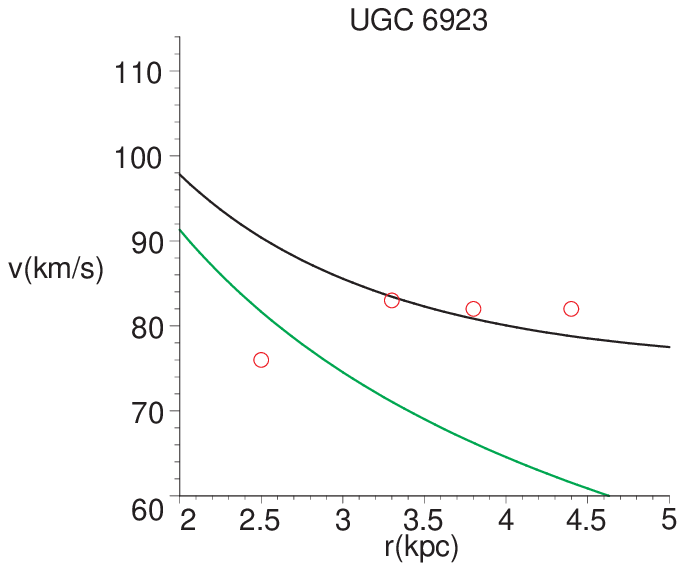}
\includegraphics[width=2.5in,height=2.5in]{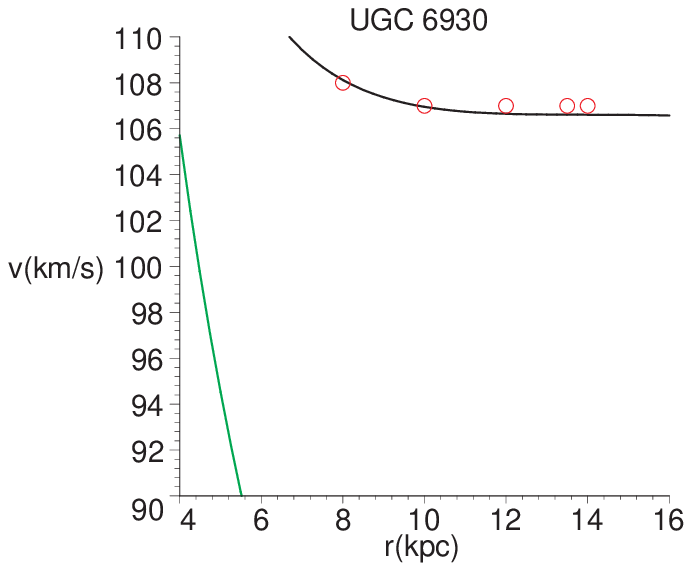}
\end{center}
\vskip 0.1 true in
\begin{center}
\includegraphics[width=2.5in,height=2.5in]{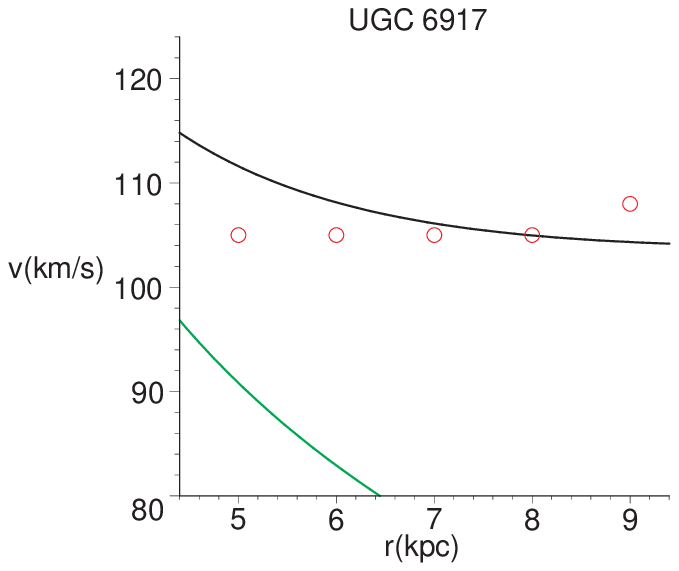}
\includegraphics[width=2.5in,height=2.5in]{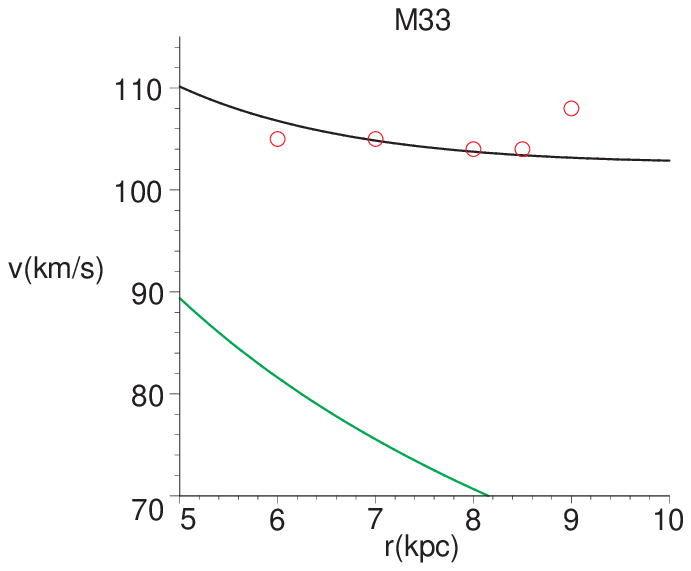}
\end{center}
\vskip 0.1 true in
\begin{center}
\includegraphics[width=2.5in,height=2.5in]{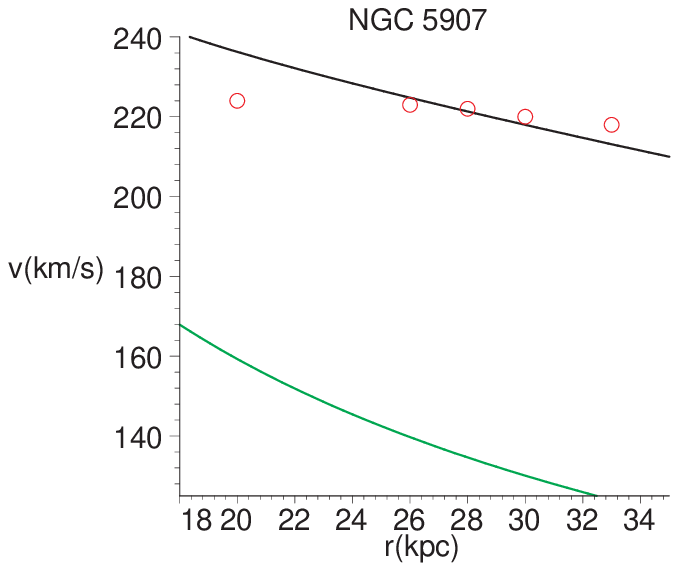}
\includegraphics[width=2.5in,height=2.5in]{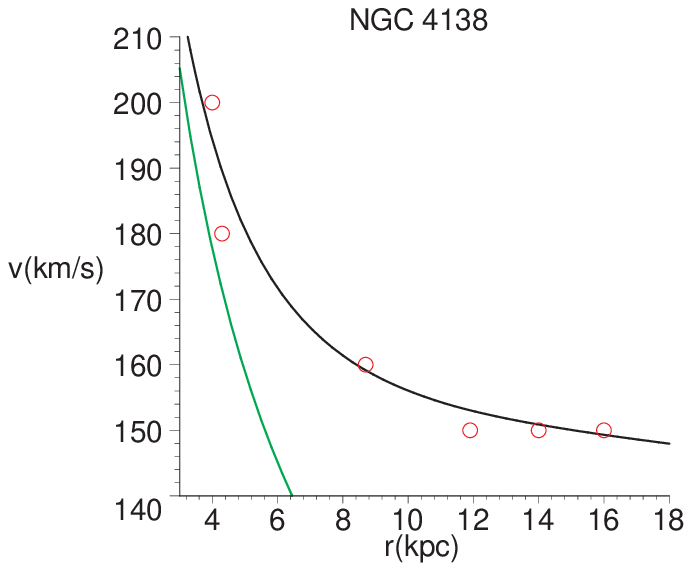}
\end{center}
\vskip 0.1 true in
\begin{center}
Fig. 1 - Fits to low-surface-brightness and
high-surface-brightness spiral galaxy data. The black curve is the
rotational velocity $v$ versus $r$ obtained from the modified
Newtonian acceleration, while the green curve shows the Newtonian
rotational velocity $v$ versus $r$. The data are shown as red
circles.
\end{center}
\vskip 0.1 true in
\begin{center}
\includegraphics[width=2.5in,height=2.5in]{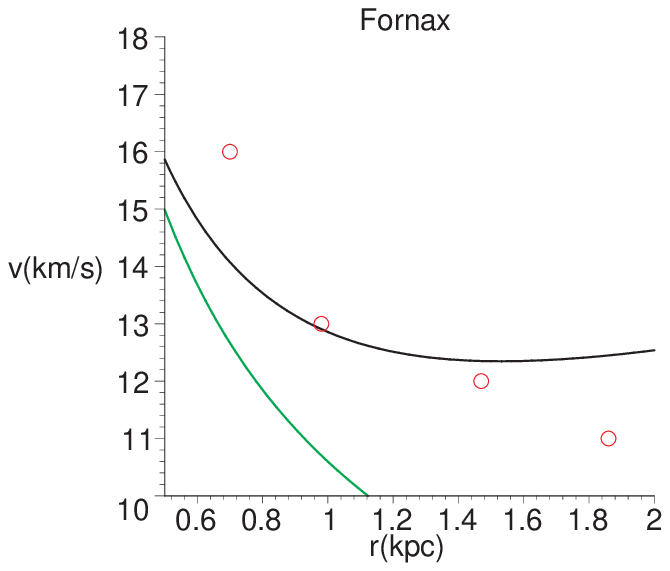}
\includegraphics[width=2.5in,height=2.5in]{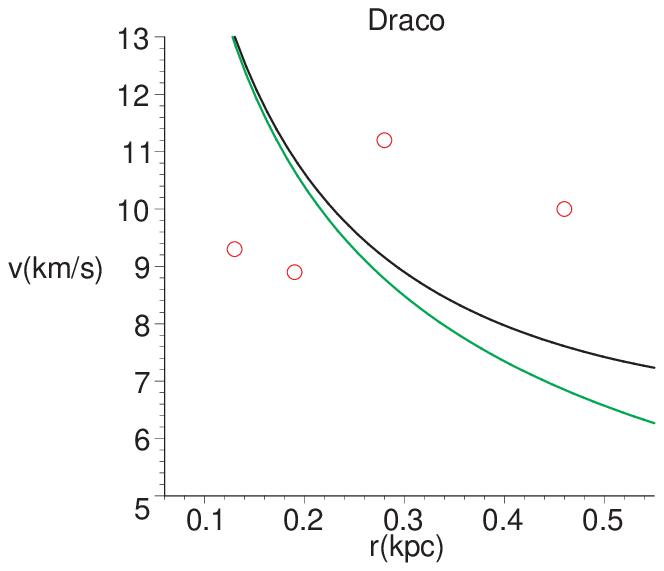}
\end{center}
\vskip 0.1 true in
\begin{center}
Fig. 2 - Fits to the data of two dwarf galaxies Fornax and Draco.
The simple relation $V\sim \sqrt{2}\sigma$ is assumed between the
velocity dispersion $\sigma$ and the rotational velocity $v$. The
black curve is the rotational velocity $v$ versus $r$ obtained
from the modified Newtonian acceleration, while the green curve
shows the Newtonian rotational velocity $v$ versus $r$. The data
are shown as red circles and the errors (not shown) are large and
for Draco the Newtonian fit cannot be distinguished from the MSTG
fit to the data within the errors.
\end{center} \vskip 0.1 true in
\begin{center}
\end{center}
\vskip 0.1 true in
\vskip 0.1 true in
\begin{center}
\includegraphics[width=2.5in,height=2.5in]{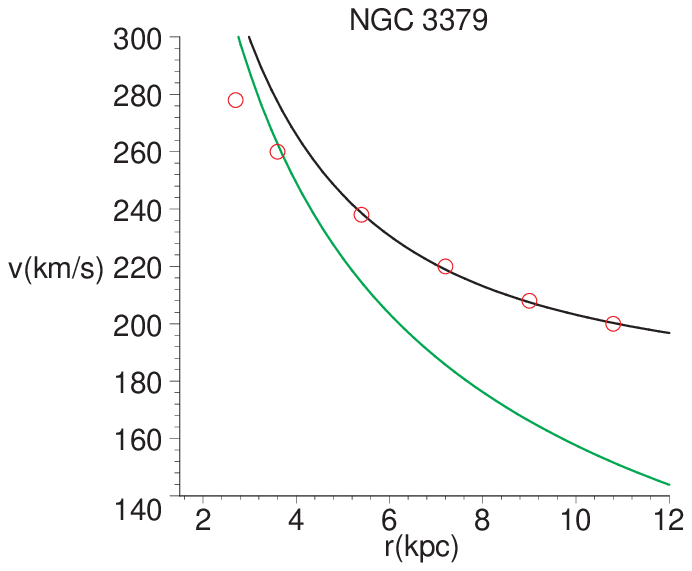}
\includegraphics[width=2.5in,height=2.5in]{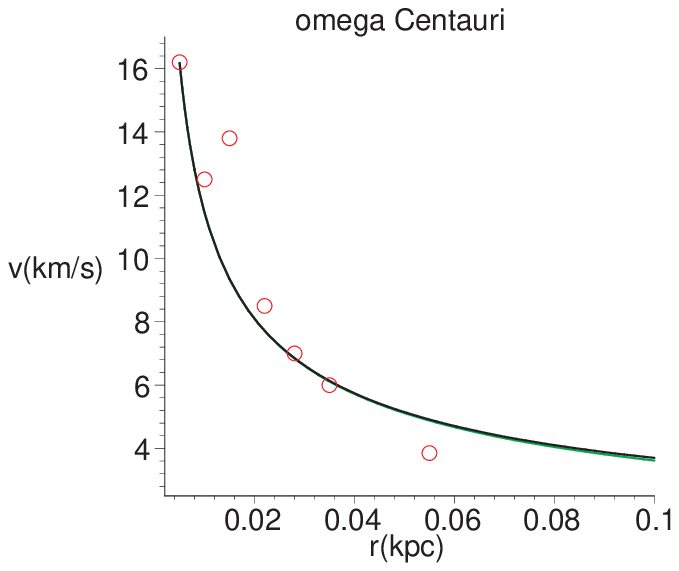}
\end{center}
\vskip 0.1 true in
\begin{center}
Fig. 3 - Fits to the elliptical galaxy NGC 3379 and the globular
cluster $\omega$ Centauri. The black curve is the rotational
velocity $v$ versus $r$ obtained from the modified MSTG
acceleration, while the green curve is the Newtonian-Kepler
velocity curve. For $\omega$ Centauri the black and green curves
cannot be distinguished from one another.
\end{center}
\vskip 0.1 true in
\begin{center}
Fig. 4 - Fits to the rotation curves for the four galaxies NGC
1560, NGC 2903, NGC 4565 and NGC 5055. The black curve is the
rotational velocity $v$ versus $r$ obtained from the modified MSTG
acceleration formula (\ref{coremodel}), while the green curve is
the Newtonian-Kepler velocity curve for the core-Newtonian
acceleration formula (\ref{coreNewton}). The luminous mass used to
fit the data for NGC 1560 is $M=1.51\times 10^{10}\,M_{\odot}$,
while the core radius is $r_c=1.5\times 10^{22}\,{\rm cm}=4.85
{\rm kpc}$. For NGC 2903 the luminous mass used to fit the data is
$M=9.95\times 10^{10}\,M_{\odot}$, while the core radius is
$r_c=2.75\,{\rm kpc}$. For NGC 4565 the luminous mass is
$M=1.81\times 10^{11}\,M_{\odot}$, while the core radius is
$r_c=3.11\,{\rm kpc}$. For NGC 5055 the luminous mass is
$M=8.55\times 10^{10}\,M_{\odot}$ and the core radius is
$r_c=2.04\,{\rm kpc}$.
\end{center}
\vskip 0.1 true in
\begin{center}
\includegraphics[width=2.5in,height=2.5in]{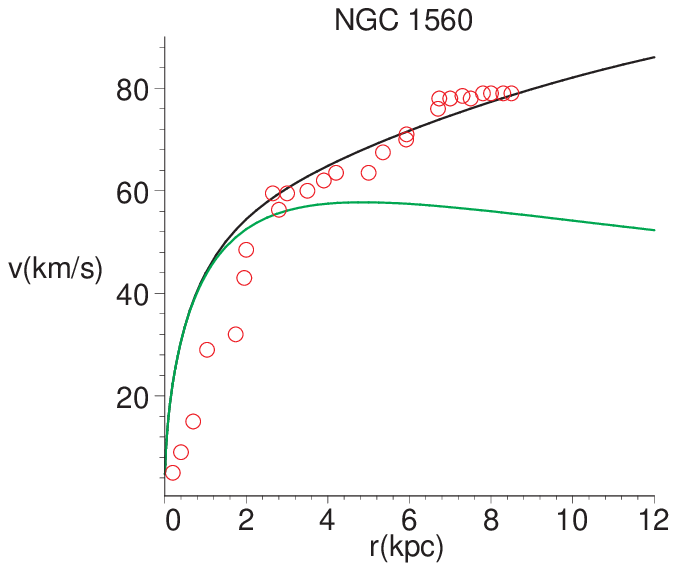}
\includegraphics[width=2.5in,height=2.5in]{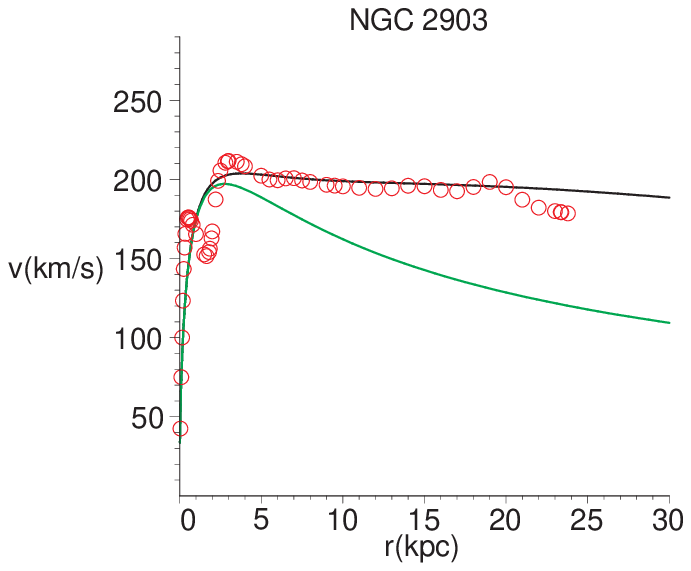}
\end{center}
\vskip 0.1 true in
\begin{center}
\includegraphics[width=2.5in,height=2.5in]{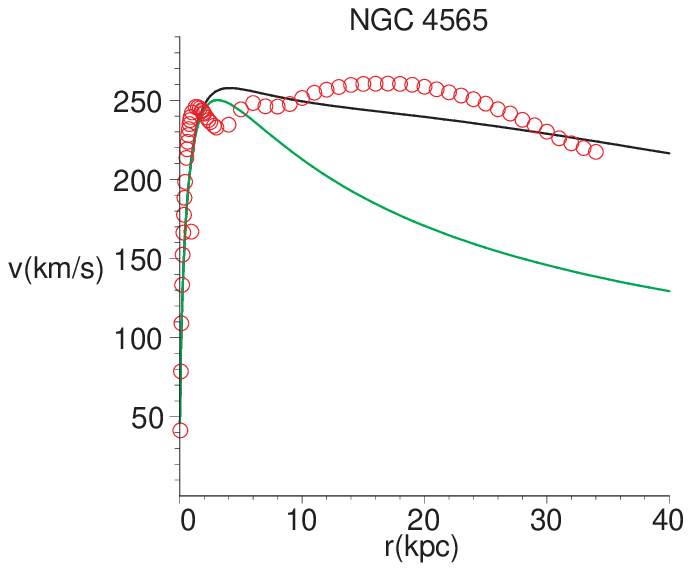}
\includegraphics[width=2.5in,height=2.5in]{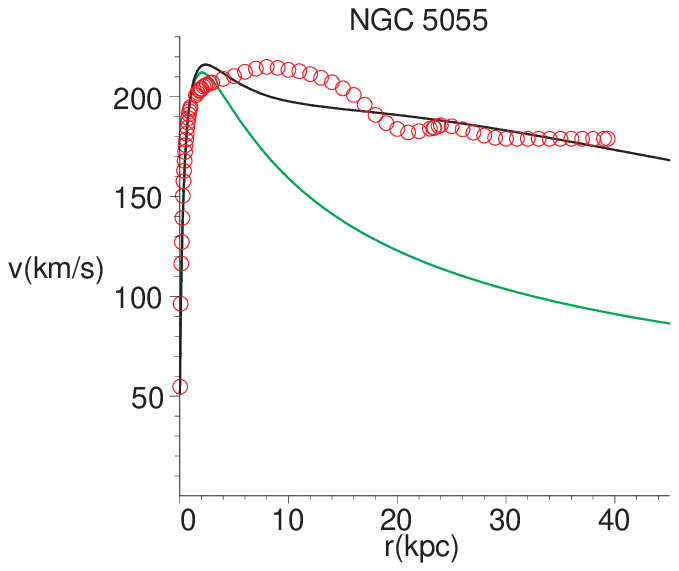}
\end{center}
\pagebreak
\begin{center}
Fig. 5 - 3-dimensional plot of $v$ versus the range of distance
$0.1\,{\rm kpc}<r<10\,{\rm kpc}$ and the range of galaxy mass
$5\times 10^6\,M_{\odot} < M < 2.5\times 10^{11}\,M_{\odot}$. The
red surface shows the Newtonian values of the rotational velocity
$v$, while the dark surface displays the MSTG prediction for $v$.
\end{center}
\vskip 0.1 true in
\begin{center}
\includegraphics[width=3.5in,height=3.5in]{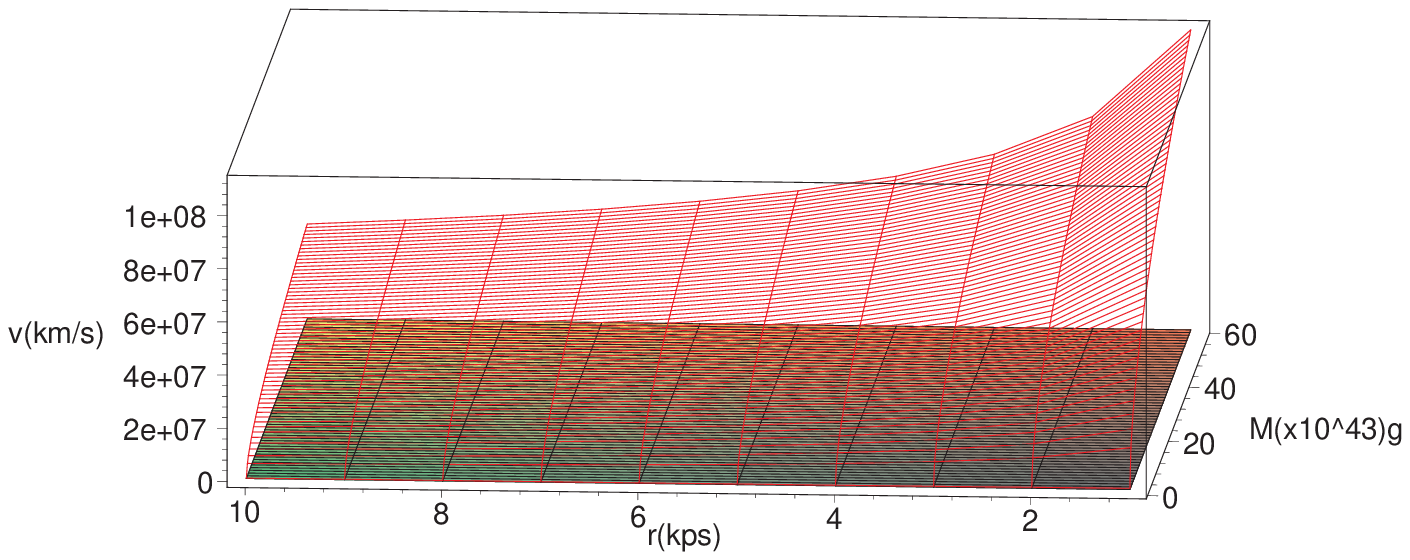}
\end{center}
\pagebreak
\begin{center}
Fig. 6 - Plot of conjectured running of the effective
gravitational constant $G(t)$ with time $t$.
\end{center}
\vskip 0.1 true in
\begin{center}
\includegraphics[width=5.5in,height=5.5in]{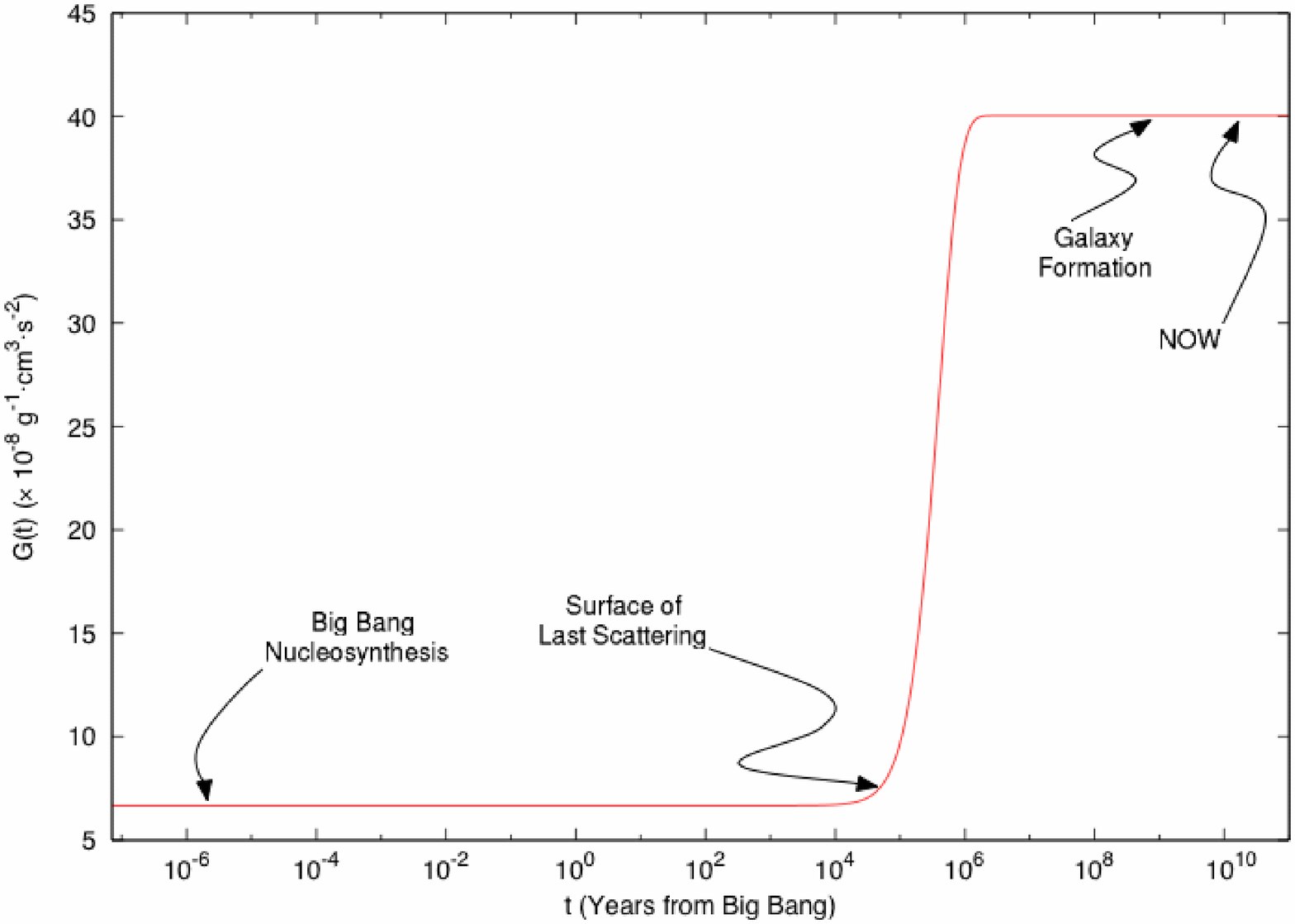}
\end{center}

\end{document}